\documentclass[aps,prb,reprint,preprintnumbers,superscriptaddress,amsmath,amssymb,bibnotes,longbibliography]{revtex4-2}

\usepackage{graphicx}
\usepackage[colorlinks,linkcolor=blue,anchorcolor=blue,citecolor=blue,urlcolor=blue,filecolor=blue,menucolor=blue,runcolor=blue]{hyperref}
\usepackage{gensymb}

\begin{document}

\title{Deformation of the triangular spin-$\frac{1}{2}$ lattice in Na$_2$SrCo(PO$_4$)$_2$}

\author{Vera P. Bader}
\email{vera.bader@uni-a.de}
\affiliation  {Experimental Physics VI, Center for Electronic Correlations and Magnetism, University of Augsburg, 86159 Augsburg, Germany}

\author{Jan Langmann}
\affiliation  {Chemical Physics and Materials Science, University of Augsburg, 86159 Augsburg, Germany}

\author{Philipp Gegenwart}
\affiliation{Experimental Physics VI, Center for Electronic Correlations and Magnetism, University of Augsburg, 86159 Augsburg, Germany}

\author{Alexander A. Tsirlin}
\email{altsirlin@gmail.com}
\affiliation{Experimental Physics VI, Center for Electronic Correlations and Magnetism, University of Augsburg, 86159 Augsburg, Germany}
\affiliation{Felix Bloch Institute for Solid-State Physics, University of Leipzig, 04103 Leipzig, Germany}

\date{\today}

\begin{abstract}

Crystal structure and thermodynamic properties of Na$_2$SrCo(PO$_4$)$_2$, the chemical sibling of the triangular quantum spin-liquid candidate Na$_2$BaCo(PO$_4$)$_2$, are reported. From single crystal x-ray diffraction and high-resolution synchrotron x-ray powder diffraction, the compound was found to crystallize in the monoclinic space group $P2_1/a$ at room temperature, in contrast to the trigonal Na$_2$BaCo(PO$_4$)$_2$. Above 650\,K, the symmetry of Na$_2$SrCo(PO$_4$)$_2$ changes to $C2/m$, while around 1025\,K a further transformation toward trigonal symmetry is observed. The monoclinic symmetry leads to a small deformation of the CoO$_6$ octahedra beyond the trigonal distortion ubiquitous in this structure type, and results in the stronger $g$-tensor anisotropy ($g_{\text{z}}/g_{\text{xy}} = 1.6 $) as well as the increased XXZ anisotropy ($J_{\text{z}}/J_{\text{xy}} = 2.1$) compared to the Ba compound ($g_{\text{z}}/g_{\text{xy}} = 1.1 $, $J_{\text{z}}/J_{\text{xy}} = 1.5$), while the average coupling strength, $J_{\text{av}}/k_{\text{B}}=(2J_{\text{xy}}+J_{\text{z}})/3k_\text{B}\simeq 1.3\,\text{K}$, remains unchanged. The N\'{e}el temperature increases from 140\,mK (Ba) to 600\,mK (Sr), and an uncompensated in-plane moment of $0.066(4)\mu_{\text{B}}/\text{f.u.}$ appears. We show that the ordering temperature of a triangular antiferromagnet is capably controlled by its structural distortions.
\end{abstract}

\maketitle


\section{Introduction}

\begin{figure*}
\includegraphics[angle=0,width=\textwidth]{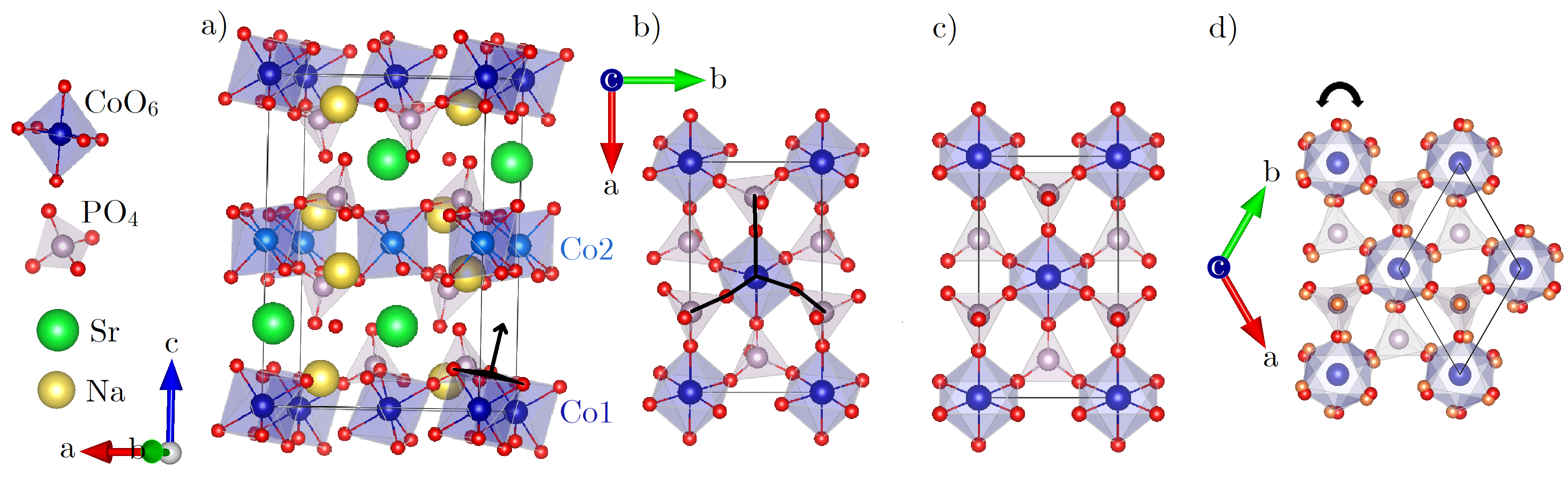}
\caption{\label{Structure} Crystal structure of Na$_2$SrCo(PO$_4$)$_2$. (a) Unit cell viewed perpendicular to the $c$ axis showing the layered structure. The tilting of the CoO$_6$ octahedra is indicated with the black arrow. (b) Triangular layer formed by the CoO$_6$ octahedra in the low-temperature monoclinic phase $P2_1/a$ viewed along the $c$ axis. The Co-O-P bonds referred to in Sec.~\ref{Struc} are drawn with thick black lines. (c) Same triangular layer in the intermediate monoclinic phase $C2/m$. (d) Triangular layer in the high-temperature trigonal phase $P\overline{3}m1$. In the top view perspective, the Na$^+$ and Sr$^{2+}$ ions were omitted for clarity.}
\end{figure*}

The search for quantum spin-liquids in triangular antiferromagnets has its beginning with Anderson's transfer of Pauling's resonating valence bond model to spin-1/2 triangular Heisenberg antiferromagnets \cite{Anderson1973}. Later it was shown theoretically and experimentally that in this nearest-neighbor (nn) isotropic Heisenberg antiferromagnet long-range magnetic order in the form of a 120\,\degree{} non collinear ground state is realized instead \cite{Huse1988, Capriotti1999}. The same holds if the model is extended from the Heisenberg to the XXZ case by introducing an anisotropy \mbox{parameter $\Delta$} \cite{Yamamoto2014}. Luckily the prospects of a quantum spin liquid state in spin-1/2 triangular antiferromagnets (TLAF) are not over here because two further adjusting screws are available to push the system into a spin-liquid state \cite{Li2020_1}. One possibility is the introduction of off-diagonal anisotropy and another one is the addition of next-nearest-neighbor (nnn) interactions $J_2$ \cite{Zhu2015,Iqbal2016,Iaconis2018}. However, it is still debated whether the off-diagonal anisotropy alone could stabilize a quantum spin liquid \cite{Luo2017, Zhang2022_1}, whereas already nnn interactions as small as 6\% of the isotropic nn interaction can push the system into a spin-liquid state \cite{Zhu2015, Zhu2018, Maksimov2019}. This offers a large playground for experimentalists who can try to tune their systems through the parameter space. Desirable are materials with undistorted triangular lattices and $\Delta$ close to 1. Enhanced XXZ anisotropy leads to a narrowing of the spin-liquid window in the above described parameter range \cite{Zhu2018} which makes it more difficult to access the spin-liquid regime.\\
\indent Existing theoretical works suggest that the spin-liquid state is unlikely to appear in a TLAF with vanishingly small $J_2$, especially if there are no pre-conditions for the large off-diagonal anisotropy. It is surprising from this perspective that the quantum spin-liquid state has been claimed in Na$_2$BaCo(PO$_4$)$_2$ based on the absence of static magnetism, as shown by the ac susceptibility measured down to 50\,mK \cite{Zhong2019} and muon spin relaxation measured down to 80\,mK \cite{Lee2021}. Na$_2$BaCo(PO$_4$)$_2$ is a $S=1/2$ TLAF. The Co$^{2+}$ ions with the effective spin-1/2 form a triangular lattice with the nn distance of around 5.3\,\AA. The magnetic layers in the $ab$ plane are stacked \mbox{along $c$} and are separated by non-magnetic ions (similar to Fig.~\ref{Structure}) \cite{Zhong2019}. Because there is no direct Co-O-Co link within the magnetic slabs, one exchange path is expected to be \mbox{Co-O-P-O-Co}, e.g., the magnetic coupling is mediated through the PO$_4$ tetrahedra \cite{Moeller2012}.
The large nnn distance of around 9.2\,\AA{} should lead to an insignificant exchange $J_2$, thus rendering Na$_2$BaCo(PO$_4$)$_2$ a purely nearest-neighbor triangular antiferromagnet, which is not expected to host a quantum spin liquid. Later indeed, the picture of a spin-liquid ground state was challenged by the phase transition observed in zero-field heat capacity at 140\,mK \cite{Li2020}, but the low-temperature properties are still controversially discussed in the literature \cite{Li2020, Wellm2021}.\\
\indent In general, quantum states of matter hinge upon a delicate balance of competing interactions that depend, in turn, on fine details of the crystal structure. For example, in Cs$_2$CuBr$_4$ the Jahn-Teller effect compresses the CuBr$_4$ tetrahedra and the spin-1/2 ions form, as a consequence, a distorted triangular lattice. This leads to a small perturbation in the form of Dzyaloshinskii-Moriya (DM) interactions, which are suppossed to be responsible for the stabilization of several plateau phases, whereas undisorted TLAFs, like Ba$_3$CoSb$_2$O$_9$, show the 1/3-plateau of the up-up-down magnetic phase only \cite{Ono2003,Fortune2009,Alicea2009,Susuki2013}. Not only the arrangement of the magnetic ions is crucial but also the non magnetic spacer ions can have a decisive influence on the properties. The 2D character of the system can be influenced by tuning the separation between the layers. Going from Ba$_3$CoNb$_2$O$_9$ to Ba$_8$CoNb$_6$O$_{24}$, that means increasing the thickness of the non-magnetic buffer layer, leads to the suppression of two subsequent phase transitions and formation of a single broad feature in the heat capacity data \cite{Rawl2017, Cui2018}.\\
\indent Here, we study the sensitivity of the disputed quantum spin-liquid state of Na$_2$BaCo(PO$_4)_2$ to structural changes induced by a chemical substitution. We show that the new compound, Na$_2$SrCo(PO$_4)_2$ obtained by replacing Ba with Sr develops an antiferromagnetic order with a sizable uncompensated magnetic moment below 600\,mK. This N\'{e}el temperature $T_N$ is more than four times higher than in the Ba case, whereas the crystal structure of Na$_2$SrCo(PO$_4)_2$ shows a clear monoclinic distortion that leads to a weak deformation of the triangular spin lattice and the enhanced easy-axis anisotropy.

         
\section{Methods}

\subsection{Sample preparation}
The growth procedure is similar to the synthesis of Na$_2$BaCo(PO$_4$)$_2$ reported in Ref.~\cite{Zhong2019}, albeit with several adjustments. Single crystals of Na$_2$SrCo(PO$_4$)$_2$ were synthesized by mixing and grinding stoichiometric amounts of Na$_2$CO$_3$ (99.999\% Sigma-Aldrich), SrCl$_2$ (anhydrous, 99.5 \% Alfa Aesar), Co(NO$_3$)$_2 \cdot$6H$_2$O (99.999\% Sigma-Aldrich), and (NH$_4$)$_2$HPO$_4$ (99\% Grüssing GmbH) together with the NaCl flux (99.99\% Alfa Aesar)  (Co:Cl 1:25) in an Ar-glovebox. The ground powder was loaded in an alumina combustion boat and placed in a tube furnace with the N$_2$ flow of 20\,sccm during the synthesis. The sample was heated to 850\,\degree C and subsequently cooled to 750\,\degree C with the rate of 1.5\,K/h. Pink crystals formed both in the boat and on the quartz tube. The amount and size of the crystals varied from one growth procedure to another, suggesting that other phases are in strong competition with Na$_2$SrCo(PO$_4$)$_2$. The crystals were removed with tweezers and/or with distilled water and selected and cut under the microscope.

\subsection{Structural characterization}
Single-crystal x-ray diffraction experiments were performed at room temperature on a translucent light pink and platelet-shaped crystal
(\textit{i})~before (dimensions 76\,$\mu$m~$\times$ 306\,$\mu$m~$\times$ 413\,$\mu$m and (\textit{ii})~after cutting with a scalpel (dimensions 76\,$\mu$m~$\times$141\,$\mu$m~$\times$ 213\,$\mu$m). The samples were attached to MiTeGen MicroMounts using perfluorinated polyether and
mounted on a BRUKER Smart-Apex diffractometer featuring a D8 goniometer in fixed-$\chi$ geometry (BRUKER). Intensity data were collected in $\omega$-scanning mode ($\Delta\omega =$
0.5\,\degree{}) using an \mbox{INCOATEC} microfocus sealed-tube x-ray source (AgK$_{\alpha}$, $\lambda$~= 0.56087\,\AA) and a BRUKER APEX II CCD-detector. Detector-to-sample distances were chosen as
40\,mm (\textit{i}) and 50\,mm (\textit{ii}).\\
\indent The as-cast sample (\textit{i}) was affected by triple twinning, while the cut sample (\textit{ii}) was a single domain. \mbox{X-ray} diffraction intensities for both samples were evaluated using the \texttt{APEX2} \cite{APEX} suite of programs and subjected to scaling and numerical absorption correction using the program \texttt{SADABS} \cite{Krause15}. Structural models were refined with the program \texttt{JANA2006}\cite{jana2006}. Further information concerning the investigated samples, the collected x-ray diffraction data and the obtained structural model for sample (\textit{ii}) are provided in Ref.~\cite{Supp}. The crystal structures were drawn in \texttt{VESTA} \cite{Vesta2011}.\\
\indent For high-resolution synchrotron x-ray powder diffraction measurements, fine powder was filled in a quartz capillary that was spun during the experiment. The powder sample was prepared by grinding single crystals.  The measurements were performed at the ID22 beamline of the ESRF (Grenoble, France) with a wavelength of 0.35423 \AA\ at different temperatures between 10 and 1025\,K using He-flow cryostat (10 to 220\,K), N$_2$-flow cryostream (150 to 450\,K), and hot-air blower (450 to 1025\,K). Structural models were refined with the program \texttt{JANA2006} \cite{jana2006}.

\begin{figure}
\includegraphics[angle=90,width=0.4\textwidth]{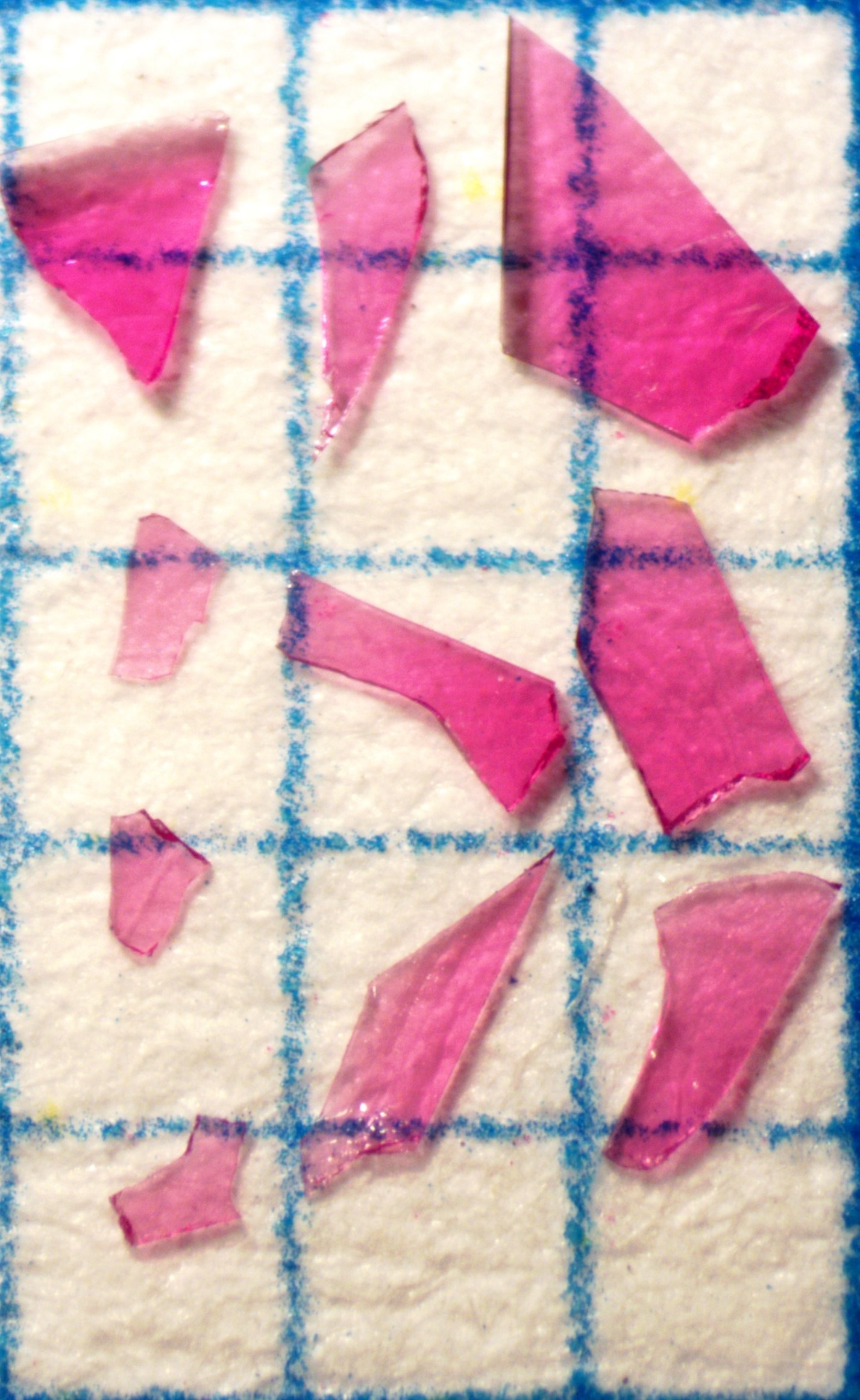}
\caption{\label{Kristalle} Pink single crystals of Na$_2$SrCo(PO$_4$)$_2$.}
\end{figure}

\subsection{Magnetization measurements}
For magnetization measurements, the MPMS3 from Quantum Design was used. The data were collected using the DC option. In-plane and out-of-plane magnetization were measured on the same 0.15\,mg single crystal. With a little amount of GE varnish the crystal was glued onto a quartz sample holder. The background contribution was negligible. In this setup, the temperature range from 2 to 400\,K was covered. Temperatures down to 400\,mK were obtained by using the $^3$He inset from Quantum Design. For this setup, the sample was glued into a plastic straw. The sample was centered at room temperature, cooled down to 400\,mK, and centered again. In both configurations magnetization was measured up to 7\,T.

\subsection{Heat capacity measurements}
The heat capacity was measured by using the relaxation method in the PPMS from Quantum Design. After the addenda measurement of platform and low-temperature grease, the 0.23\,mg single crystal was glued onto the platform. With the $^3$He option, measurements down to around 0.4\,K were possible. Due to the light weight of the sample, the upper temperature was limited to 10\,K. The data were collected for different fields ranging from 0 to 5\,T.

\subsection{\textit{Ab initio} calculations}
Density-functional-theory (DFT) band-structure calculations were performed in the \texttt{FPLO}~\cite{fplo} code using the Perdew-Burke-Ernzerhof (PBE) approximation for the exchange-correlation potential~\cite{pbe96}. Lattice parameters and atomic positions for Na$_2$SrCo(PO$_4)_2$ were taken from the structure refinement against the single-crystal XRD data. The PBE calculations were used to assess orbital energies of Co$^{2+}$. Additionally, full-relativistic DFT+$U$ was used to compute total energies of collinear spin configurations and obtain exchange couplings by a mapping procedure assuming $S=\frac32$ for Co$^{2+}$. The on-site Coulomb repulsion and Hund's coupling of DFT+$U$ were chosen as $U_d=5$\,eV and $J_d=1$\,eV, respectively~\cite{Wellm2021}.

\section{Results}
\subsection{Room-temperature crystal structure}
\begin{figure}
\includegraphics[angle=0,width=0.49\textwidth]{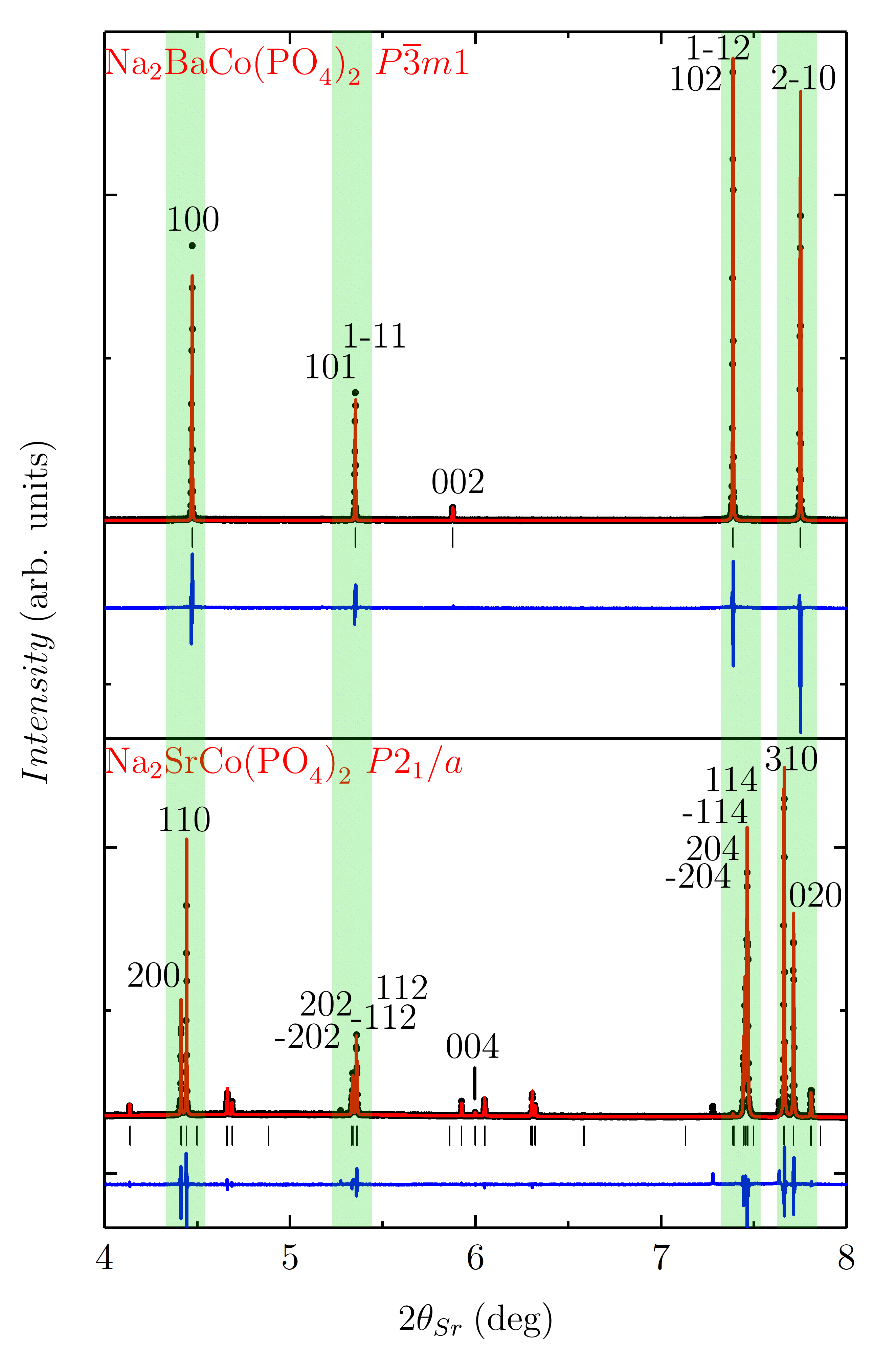}
\caption{\label{Refinement300K} Rietveld refinement for the synchrotron XRD data collected at 300\,K in the trigonal space group for Na$_2$BaCo(PO$_4$)$_2$ \cite{Bader_unpublished} and in the monoclinic space group $P2_1/a$ for Na$_2$SrCo(PO$_4$)$_2$. The illustrated peak splitting is indicative of the monoclinic symmetry of the Sr compound. The scattering angles for Na$_2$BaCo(PO$_4$)$_2$ were renormalized for an easier comparison with the Sr compound.}
\end{figure}


The pink single crystals of Na$_2$SrCo(PO$_4$)$_2$ are shown in Fig.~\ref{Kristalle}. The crystals are thin platelets, the largest ones having an area of approximately 1\,mm$^2$. In contrast to trigonal Na$_2$BaCo(PO$_4$)$_2$, at room temperature Na$_2$SrCo(PO$_4$)$_2$ crystallizes in the monoclinic symmetry, as evident from the peak splittings in the high-resolution XRD data shown in Fig.~\ref{Refinement300K} \cite{Bader2022}.
The initial model was derived from the Na$_2$BaCo(PO$_4$)$_2$ structure \cite{Zhong2019} by transforming the basis vectors as follows: $\vec{a}\,'=\vec{a}-\vec{b}$, $\vec{b}\,'=\vec{a}+\vec{b}$, and $\vec{c}\,'=2\vec{c}$. Using powder x-ray diffraction (PXRD) data the lattice constants at room temperature were refined to be $a=9.20152(2)$\,$\mathrm{\AA}$, $b=5.26593(1)$\,$\mathrm{\AA}$, $c=13.54116(2)$\,$\mathrm{\AA}$, and $\beta= 90.06613(12)$\,\degree .\\
\indent Extinction conditions derived from powder and single-crystal x-ray diffraction (SCXRD) suggest the choice of a primitive space group and indicate the presence of a $2_1$ screw axis $\parallel \vec{b}$ and an $a$ glide plane. Furthermore, comparison of structural refinements (PXRD and SCXRD) using the two space groups ($P 2/a$, $P 2_1/a$) leads to reasonable atomic parameters for $P 2_1/a$ (No. 14) only. The positions of allowed and forbidden reflections in reconstructed reciprocal-space planes from SCXRD data and the parameters obtained from the Rietveld refinement of the data collected at 300\,K with the space group $P2_1/a$ are listed in Ref.~\cite{Supp}.\\
\indent The unit cell of Na$_2$SrCo(PO$_4$)$_2$ is shown in Figs.~\ref{Structure}(a) and (b).
CoO$_6$ octahedra form planar triangular layers, which are stacked along the $c$ axis. The CoO$_6$ octahedra in the slabs are bridged by PO$_4$ tetrahedra. Na$^+$ ions fill the voids in the layers and the Sr$^{2+}$ ions act as spacer ions and create a buffer layer, which separates the magnetic layers. These structural considerations hold for both Na$_2$SrCo(PO$_4$)$_2$ and Na$_2$BaCo(PO$_4$)$_2$. In the latter case the larger Ba$^{2+}$ ions separate the Co layers. As expected for the Sr case, the magnetic layers are closer to each other, with the Co-Co distance of 6.77\,\AA{} along $c$ compared to 7.01\,\AA{} for the Ba compound at 300\,K. We note that with respect to the starting Na$_2$BaCo(PO$_4$)$_2$ structure from Ref.~\cite{Zhong2019} an additional origin shift of $(0~0~0.5)$ was necessary for Na$_2$SrCo(PO$_4$)$_2$. In this way, the Co1 atom is positioned at the origin $(0~0~0)$ of the unit cell.\\
\indent One key difference that leads to the lower monoclinic symmetry of the Sr compound compared to the trigonal symmetry of the Ba compound is the cooperative tilting of the CoO$_6$ octahedra and PO$_4$ tetrahedra. Different tilting in two adjacent layers causes the doubling of the unit cell along $c$.\\
\indent Another difference is that the symmetry lowering allows distortions of the CoO$_6$ octahedra with the local symmetry of the Co$^{2+}$ ion reduced from $D_{3d}$ 
to $C_i$. The trigonal distortion, captured by taking the O\hbox{-}Co\hbox{-}O angle with the largest deviation from 90\,\degree{}, is 4.8(2)\% in the Ba compound and amounts to a maximum deviation of 6.0(3)\% and 7.9(3)\% for the Co1O$_6$ and Co2O$_6$ octahedra in the Sr compound, respectively. Additionally, symmetry lowering leads to three different Co-O lengths in the \mbox{Sr compound} compared to only one in the Ba compound. This distortion beyond trigonal, i.e. the maximum deviation from the mean distance of the three Co-O bonds ($d_1$, $d_2$, and $d_3$, see Fig.~\ref{fig:DFT}) is 1.5(7)\% in the Co1O$_6$ octahedra and 2.2(7)\% in the Co2O$_6$ octahedra. For the PO$_4$ tetrahedra the symmetry reduces from $C_{3v}$ 
to $C_1$. In the Ba compound, the Co$^{2+}$ ions form a triangular lattice with equilateral triangles, whereas in the distorted Sr compound, the triangles are not equilateral but isosceles with one Co-Co distance being around 0.7\% shorter compared to the two other distances at room temperature. 

\subsection{Structural phase transitions}
\label{Struc}
\begin{figure*}
\includegraphics[angle=0,width=\textwidth]{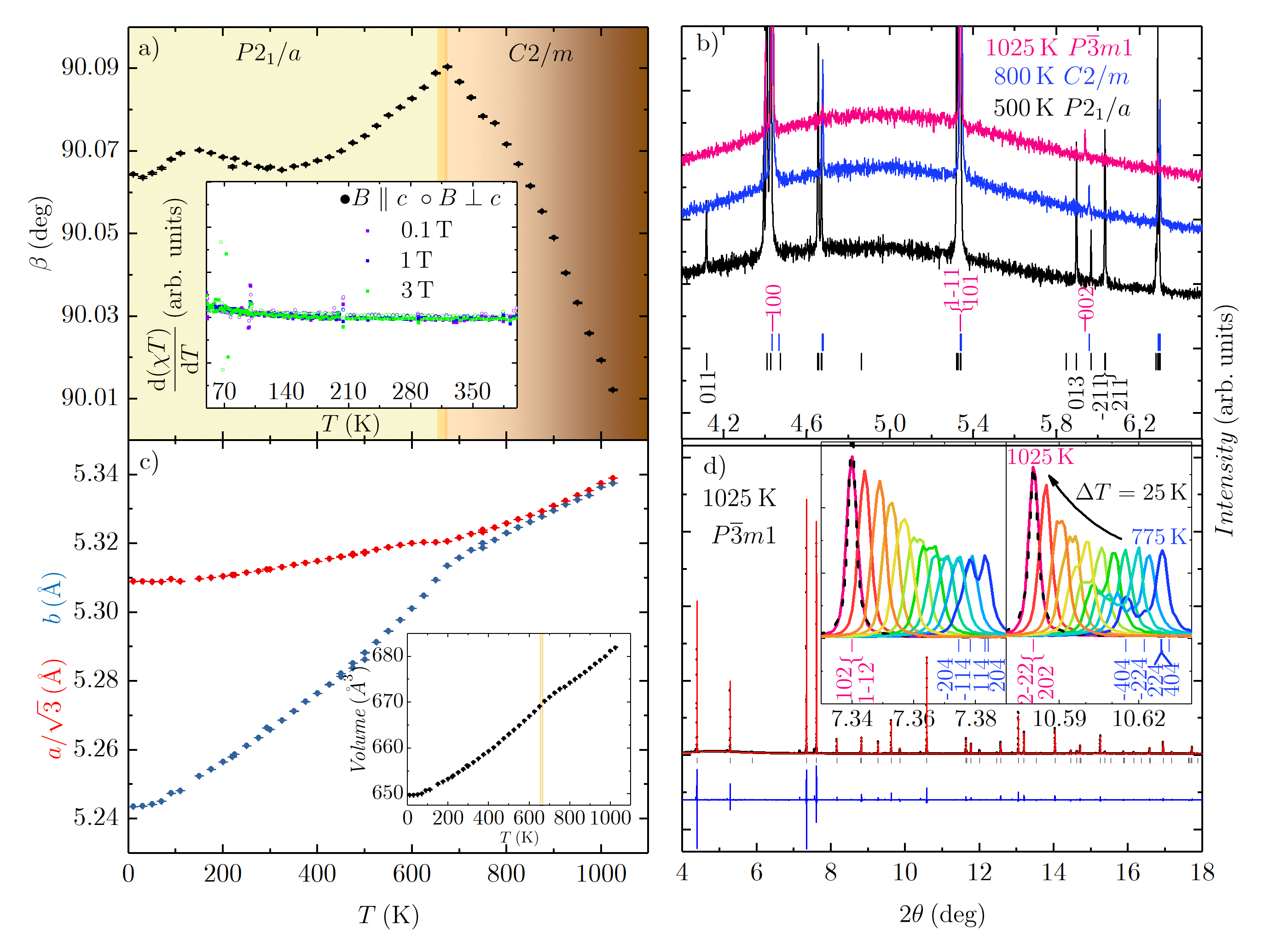}
\caption{\label{refinements} Crystal structure of Na$_2$SrCo(PO$_4$)$_2$ as a function of temperature. (a) Temperature dependence of the monoclinic angle $\beta$. The weak anomaly at 200\,K is likely a measurement artifact, because thermodynamic measurements, including Fisher's heat capacity $\text{d}(\chi T)/\text{d}T$ (inset), reveal no anomaly at this temperature. The second anomaly at around 650\,K marks the structural transition from the space group $P2_1/a$ to $C2/m$. The color gradient marks the gradual transition to a trigonal space group which is not fully completed at 1025\,K. (b) Synchrotron XRD data for selected temperatures illustrating the symmetry changes. The reflections labeled in black vanish for $C2/m$ and the red labels indicate the remaining reflections for $P\overline{3}m1$. The scattering angles were rescaled with respect to the data collected at 500\,K. (c) Red are the refined $a/\sqrt(3)$ and blue are the refined $b$ lattice parameters. Visible is the kink at the phase transition from $P2_1/a$ to $C2/m$ at around 650\,K in both lattice parameters. At higher temperatures $a/\sqrt(3)$ and $b$ converge due to the transition into a trigonal structure. The inset shows the continuous change of the unit cell volume with temperature. (d) Rietveld refinement for the synchrotron XRD  data collected at 1025\,K in the trigonal space group $P\overline{3}m1$. The insets show the merging of the split peaks as the temperature increases.}
\end{figure*}

\begin{figure*}
\includegraphics[angle=0,width=\textwidth]{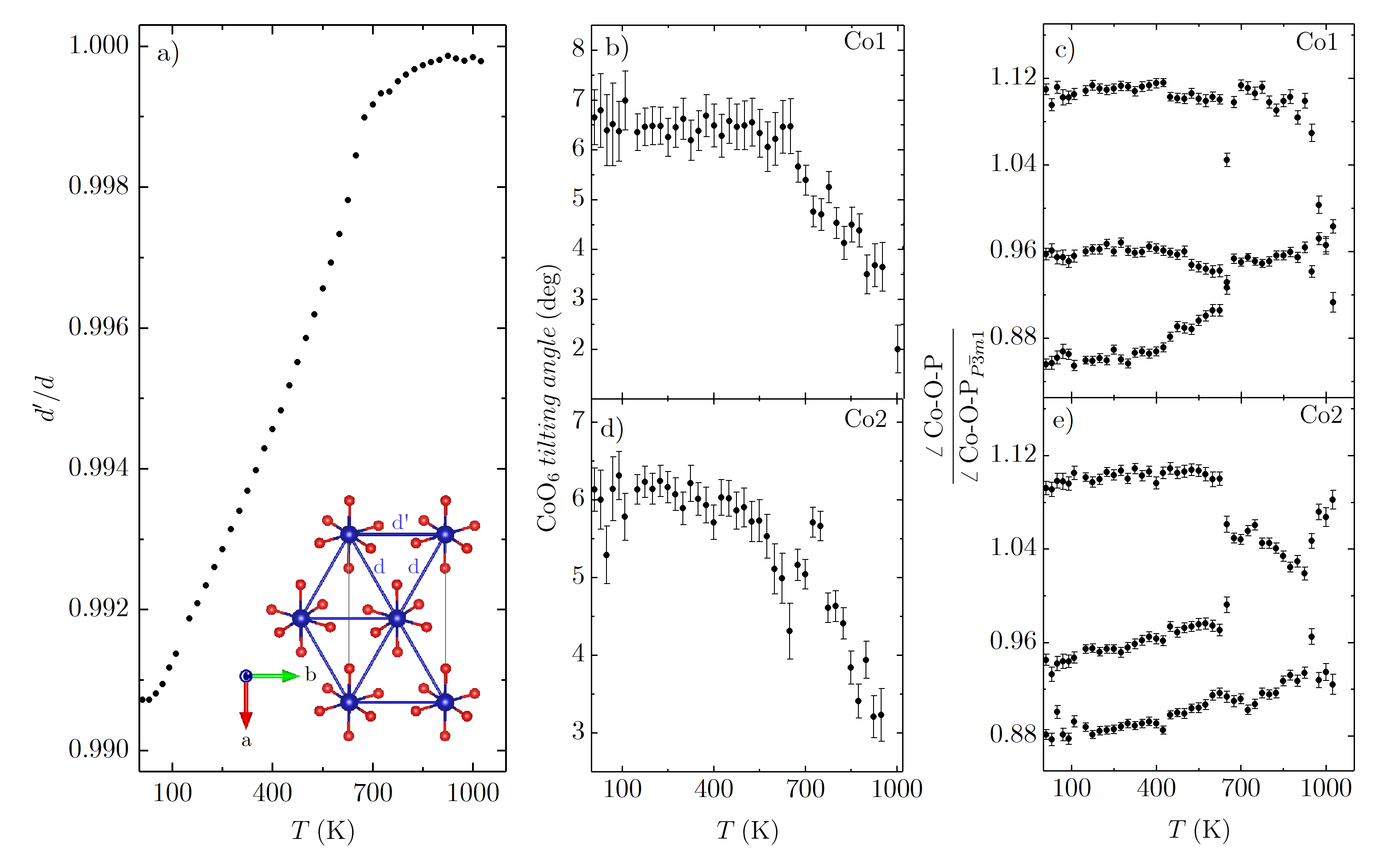}
\caption{\label{distortion} Deviation from the trigonal structure. (a) The ratio of short ($d^{\prime}$) to long ($d$) Co-Co bonds gauges deformation of the triangular lattice. The error bars are smaller than the symbol size. [(b) and (d)] Tilting angles of the Co1O$_6$ and Co2O$_6$ octahedra, respectively. [(c) and (e)] Co1-O-P and Co2-O-P angles, respectively. The angles were normalized in respect to the angle in the trigonal case.}
\end{figure*}
\begin{figure}
\includegraphics[angle=0,width=0.48\textwidth]{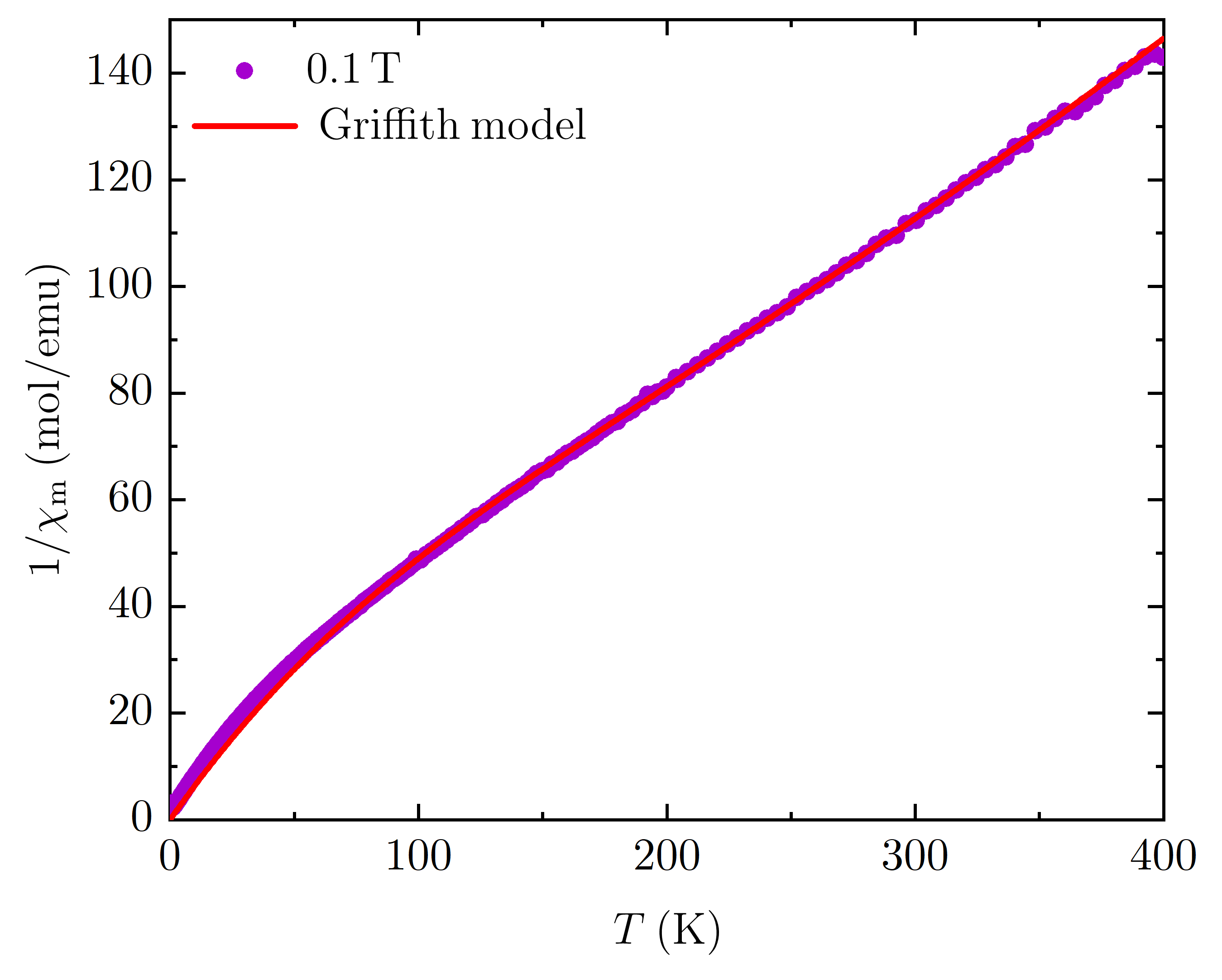}
\caption{\label{Griffith} Inverse magnetic susceptibility data at 0.1\,T. The red line depicts the fit with the Griffith model as described in Ref.~\cite{Griffith1958}.}
\end{figure}

In this section, we investigate the temperature dependence of the crystal structure because several aspects indicate a structural phase transition. Since Na$_2$SrCo(PO$_4$)$_2$ is a derivative of Na$_2$BaCo(PO$_4$)$_2$, the distortion may be reverted upon heating. In addition, the triple twinning of the crystal structure seen in the SCXRD data indicates that Na$_2$SrCo(PO$_4$)$_2$ grows in a higher symmetry and then transforms into mono\-clinic upon cooling. A structural change from monoclinic to trigonal was already reported for different glaserite-type compounds \cite{Sanjeewa2017, Sanjeewa2019, Reuss2018, Sanjeewa2016, Tapp2017}.\\
\indent To probe structural phase transitions, the synchrotron x-ray patterns 
at different temperatures between 10 and 1025\,K were collected \cite{Bader2022, Bader2022_1}. No anomalies were observed below 650\,K, except for a small peak in the monoclinic angle $\beta$ at 200\,K [Fig.~\ref{refinements}(a)]. Since none of the $a$, $b$, $c$ lattice parameters show any anomaly at this temperature [Fig.~\ref{refinements}(c)], nor any feature is visible in the susceptibility and Fisher's heat capacity [Figs.~\ref{Griffith} and \ref{refinements}(a)], we believe that this small peak may be a measurement artifact caused by the change in the sample environment. On the other hand, prominent changes take place at 650\,K. Above this temperature, the symmetry remains monoclinic, but some of the reflections disappear. The comparison of the data collected at 500 and at 800\,K is shown in Fig. \ref{refinements}(b). The reflections labeled in black vanish above 650\,K. The new extinction condition $h+k=2n$ is indicative of the $C$ centering. A continuous evolution of the unit-cell volume [inset of Fig.~\ref{refinements}(c)] further suggests that the transition is of second order and should thus follow the group-subgroup relation. From all these considerations, the space group with the highest possible symmetry is $C2/m$. The top view of the unit cell is shown in Fig.~\ref{Structure}(c). In $P2_1/a$ the tilting angle is the same for the CoO$_6$ octahedra in one layer but opposite components of the tilting vector in the $b$ direction forbid the vector \mbox{(0.5 0.5 0)} as a translational symmetry element. In contrast, in $C2/m$ the $b$ component vanishes, confining the tilting to the $ac$ plane. The octahedra within the layers tilt the same way and the unit cell becomes $C$-centered. The $c$ doubling is still present in the structure. \mbox{Rietveld} refinement of the exemplarily chosen data at 800\,K in the space group $C2/m$ describes the data accurately and is shown, together with the refined atomic parameters, in the Supplemental Material \cite{Supp}. The observed phase transformation to $C2/m$ was reversible and the room-temperature phase $P2_1/a$ was recovered upon cooling.\\
\indent On further increasing temperature, the values of the $a/\sqrt(3)$ and $b$ lattice parameters converge. The $a/\sqrt(3)$ and $b$ converging and the monoclinic angle $\beta$ approaching 90\,\degree , both indicate the transition to a trigonal symmetry. The $C2/m$ space group can be regarded as intermediate step in the $P2_1/a$ to $P\overline{3}m1$ transition series \cite{Boukhris2012}. This is illustrated in Fig. \ref{refinements}(b), in which the red labels indicate the remaining reflections for $P\overline{3}m1$. That symmetry constrains $a/\sqrt(3)$ and $b$ to be equal and restricts $\beta$ to 90\,\degree{} becomes clear from the fact that many split peaks merge. This is shown in the inset of Fig.~\ref{refinements}(d) for selected peaks. The clearly separated peaks at 775\,K merge as the temperature is increased until the splitting cannot be resolved anymore at around 1000\,K. At 1025\,K, $\beta$ has not reached 90\,\degree{} but equals  90.0122(2)\,\degree . The transition is not yet completed at 1025\,K but the data can be also refined in the space group $P\overline{3}m1$ by taking into account residual broadening via the increased reflection width. In this way, the diffraction pattern can be described accurately with a residue value $R_{\text{p}}=9.36\%$ [Fig.~\ref{refinements}(d)], which is only slightly higher than the residue value of the refinement in the monoclinic structure ($R_{\text{p}}=8.59\%$). The Lorentzian parameter of the peak-shape function $LY$ has the value of 0.0412(2)\,\degree{} for the refinement in the trigonal space group compared to $LY=0.0352(2)$\,\degree{} for the refinement in the monoclinic model. The atomic parameters in the trigonal space group are listed in Ref.~\cite{Supp}. We believe that at 1025\,K, Na$_2$SrCo(PO$_4)_2$ is on the verge of a phase transition into the trigonal structure. Unfortunately, technical reasons prevented us from performing measurements at even higher temperatures where peak broadening due to the minute monoclinic distortion could be fully suppressed.\\
\indent The results from room-temperature SCXRD experiments also hint at the proximity of monoclinic Na$_2$SrCo(PO$_4$)$_2$ to the trigonal aristotype structure with the space group $P\overline{3}m1$: An as-cast sample was subject to growth twinning with three monoclinic domains (space group $P2_1/a$; domain fractions 87.5(3)\%~: 6.6(2)\%~: 5.9(3)\%) and twin element $\overline{3}$ oriented parallel to the $c$ axis of the trigonal unit cell (see Ref.~\cite{Supp}). Cutting with a scalpel resulted in a single-domain sample underlining the macroscopic nature of the twin domains.\\
\indent The $P\overline{3}m1$ unit cell is shown in Fig.~\ref{Structure}(d). As described above the structure is undistorted and the polyhedra are not tilted. In the Sr compound ferrorotations of the CoO$_6$ octahedra can be easily introduced in the magnetic layers which is illustrated in Fig.~\ref{Structure}. The x rays record the average $P\overline{3}m1$ structure.\\
\indent The distortions appearing on cooling can be quantified by the ratio of the short ($d^{\prime}$) and long ($d$) nearest-neighbor Co-Co distances. This ratio is depicted in Fig.~\ref{distortion}(a). With increasing temperature the difference in length decreases  and the structural phase transition at 650\,K is apparent from the change of the slope. In $C2/m$ the distortions still forbid trigonal symmetry but the triangles are nearly equilateral and the ratio approaches 1.0 which corresponds to the undistorted lattice in the $P\overline3$m1 case.\\
\indent The tilting of the CoO$_6$ octahedra, indicated in Fig.~\ref{Structure}(a) and shown in Figs.~\ref{distortion}(b) and \ref{distortion}(d), is quantified by the angle between the normal of the indicated face of the octahedron and the $c$ axis. In the untilted case ($P\overline{3}m1$), the angle is 0\,\degree{}. In the space group $P2_1/a$, the octahedra are tilted the most, and in both Co layers the tilting angle stays nearly constant with temperature. After the phase transition to $C2/m$ at around 650\,K, the angles decrease until they reach at 1000\,K around 3\,\degree{}, which is half the starting value.\\
\indent The tilts of both CoO$_6$ octahedra and PO$_4$ tetrahedra affect the Co-O-P angles, which are visualized in Fig.~\ref{Structure}(b). The Co-O-P unit is part of the magnetic exchange path and the temperature dependent Co-O-P bond angles are shown in Figs.~\ref{distortion}(c) and \ref{distortion}(e). In the untilted case ($P\overline{3}$m1) all the angles are the same and the depicted angles for different temperatures are normalized in regard to that value obtained at 1025\,K. A larger deviation from the reference value is observed at the lowest temperatures. The three different angles indicate that the superexchange paths are different. With increasing temperature the spread of the angles decreases until two angles in each Co layer merge at the $C2/m$ phase transition. The two different angles would finally coincide when the transition to $P\overline{3}m1$ is completed.

\begin{figure}
\includegraphics[angle=0,width=0.46\textwidth]{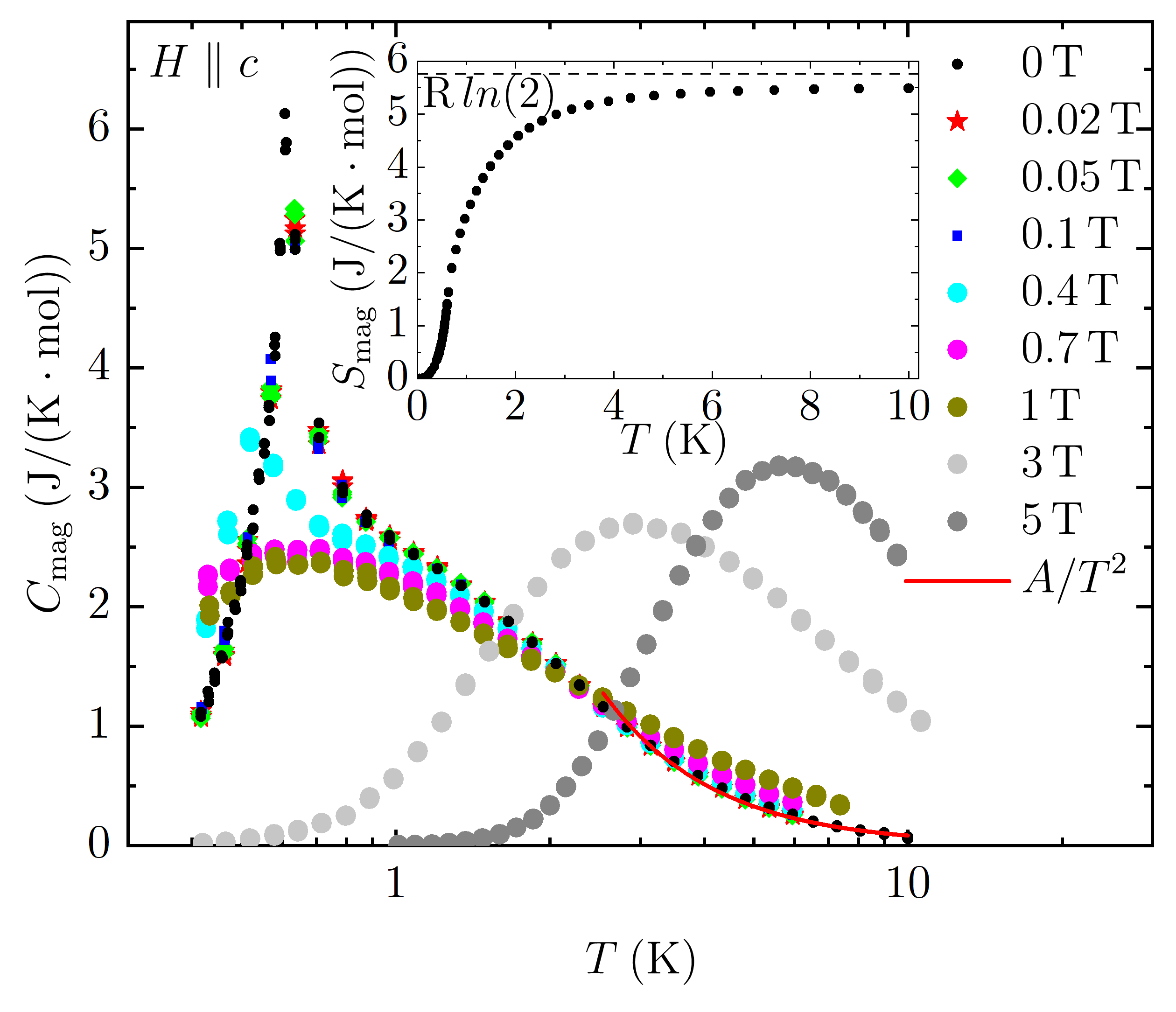}
\caption{\label{PPMS} Temperature dependence of the magnetic heat capacity at different magnetic fields parallel to $c$. The broad Schottky-like peak shifts to higher temperatures with increasing magnetic field. In low magnetic fields the broad peak due to short-range order is superimposed on the $\lambda$-type anomaly that indicates the magnetic ordering transition at 600\,mK. The fit of the zero-field data is depicted as red line. The inset shows the zero-field magnetic entropy, which approaches $\textrm{R}ln(2)$.}
\end{figure}

\subsection{Heat capacity}
\begin{figure}
\includegraphics[angle=0,width=0.48\textwidth]{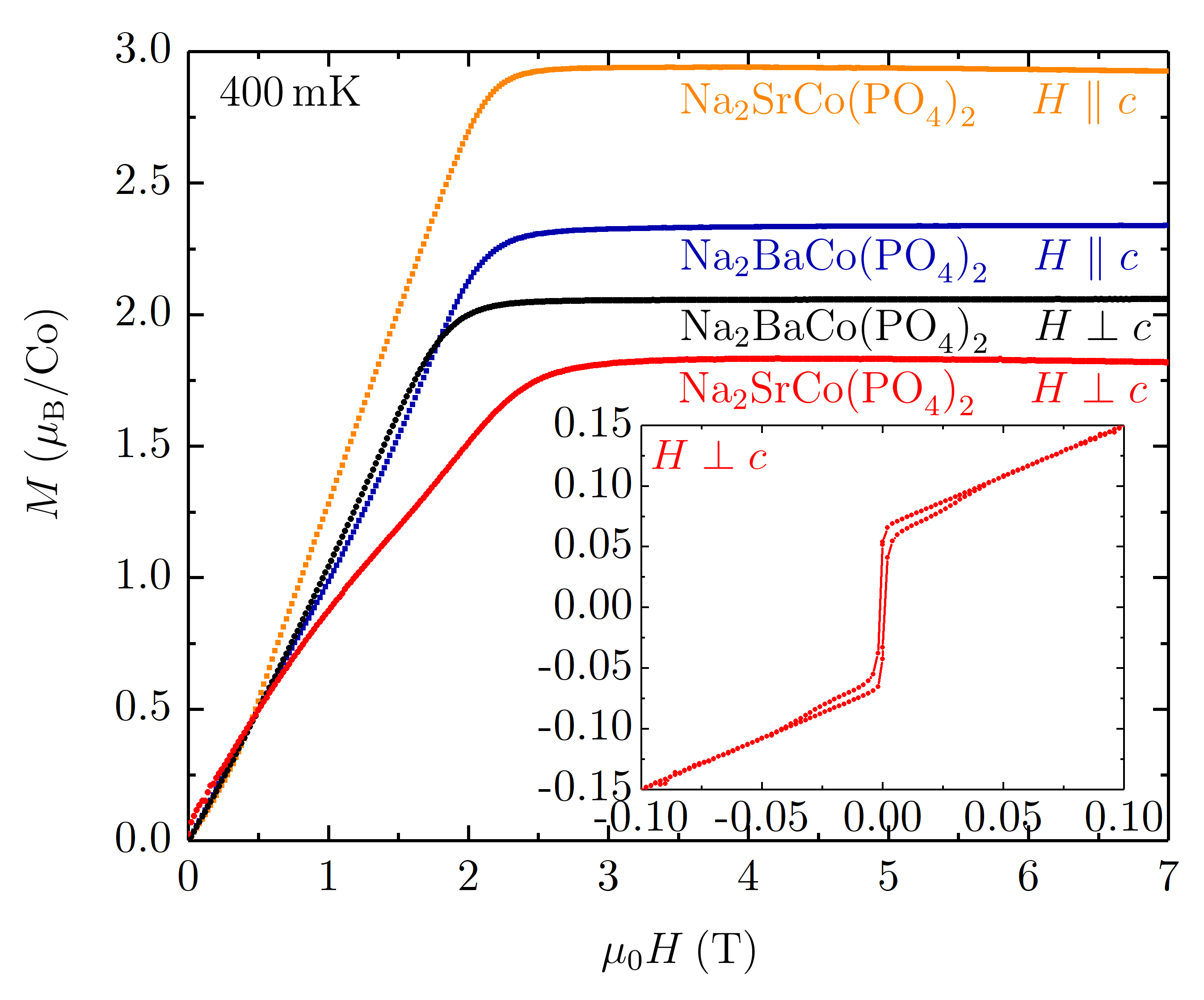}
\caption{\label{MvH} Comparison of the magnetization versus magnetic field for Na$_2$SrCo(PO$_4$)$_2$ and Na$_2$BaCo(PO$_4$)$_2$ \cite{Bader_unpublished} for both field directions at 400\,mK. The Sr compound shows a three times larger magnetic anisotropy. The inset shows the small hysteresis that indicates the uncompensated spin component in the $ab$ plane.}
\end{figure}

The magnetic specific heat (after phonon subtraction) measured on a single crystal in different magnetic fields is shown in Fig.~\ref{PPMS}. In zero field, a sharp peak at around $T_N\simeq 600$\,mK indicates long-range magnetic order. Additionally, a hump centered around 1\,K indicates short-range magnetic order above $T_N$, as typical for low-dimensional and frustrated antiferromagnets. On increasing the field, $T_N$ initially does not change and then decreases to 520\,mK at 0.4\,T. In higher fields, only a broad maximum remains, while $T_N$ is suppressed below 0.4\,K or vanishes. This behavior is broadly consistent with theoretical results for classical triangular antiferromagnets \cite{Gvozdikova2011} and with experimental results obtained on model compounds, such as Ba$_3$CoSb$_2$O$_9$ \cite{Zelenskiy2019}. One noteworthy feature is that $T_N$ drops by 13\% at only 20\% of the saturation field, i.e., it is suppressed faster than in other triangular antiferromagnets.\\
\indent In fields above 1\,T, the broad maximum systematically shifts toward higher temperatures and resembles a \mbox{Schottky} anomaly, suggesting that a fully polarized state has been reached, and the field-induced gap opens. Indeed, our magnetization data suggest saturation around 2\,T (Fig.~\ref{MvH}).\\
\indent The phonon contribution was extracted by fitting the zero-field data between 2.5 to 10\,K to $B\cdot T^3 + A/T^2$ (see Ref.~\cite{Supp} and Fig.~\ref{PPMS}), where the first term is the phonon part and the second one is the lowest-order term in the high-temperature series expansion (HTSE) for the magnetic specific heat \cite{Johnston2000} ($B=2.6(2)\cdot10^{-4}\,\textrm{J}/\textrm{(K}^4\cdot\textrm{mole)}$, $A=8.3(1)\,(\textrm{J}\cdot \textrm{K} /\textrm{mole})$). In the nearest-neighbor triangular antiferromagnet, $A=(9/16)\cdot R \cdot J^2$ and $J/k_{\text{B}}=1.33(1)\,\text{K}$ is obtained. We will see that this value matches well to those determined in Sec.~\ref{Magn_Results}.\\
\indent Before calculating the entropy by integration, the zero field data were extrapolated to 0\,K using the proportionality $C \propto T^3$ assuming a gapless antiferromagnet. The entropy is shown in the inset of Fig.~\ref{PPMS} and approaches the value $\textrm{R}\,ln(2)$ expected for a spin-1/2 system. The small deviation of 5\% may be caused by the weighing error for the small single crystal. This indicates that below 10\,K, Na$_2$SrCo(PO$_4$)$_2$ features effective spin-1/2 magnetic moments, similar to Na$_2$BaCo(PO$_4$)$_2$.

\begin{figure*}
\includegraphics[angle=0,width=0.9\textwidth]{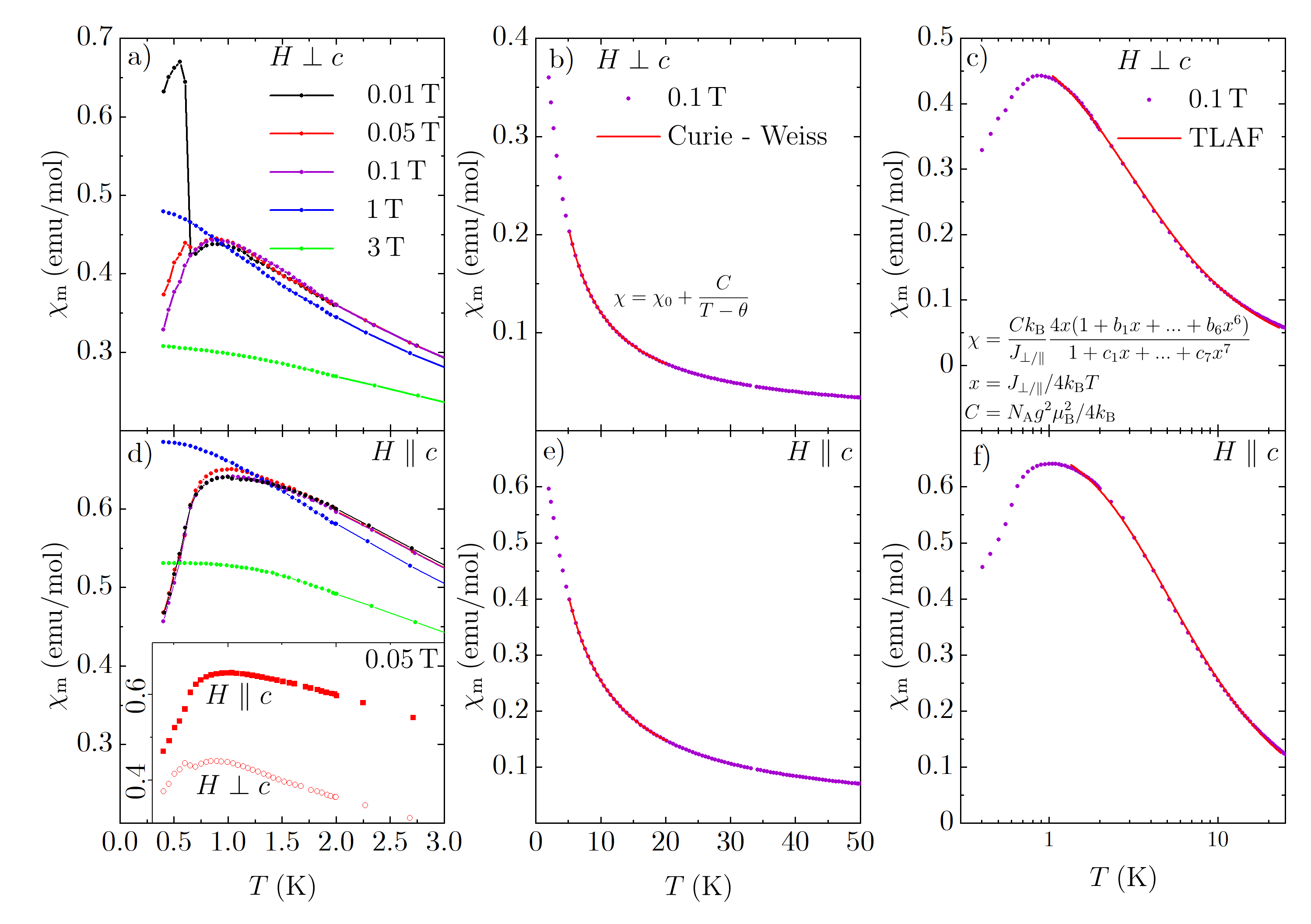}
\caption{\label{Magn} (Top) Magnetic field perpendicular to $c$. (Bottom) Magnetic field parallel to $c$. [(a) and (d)] Magnetic susceptibility for different magnetic fields below 3\,K. The antiferromagnetic transition manifests itself by the drop of the susceptibility and remains visible up to at least 0.1\,T. The inset contrasts the susceptibilities for both field directions at 0.05\,T. In low fields up to 0.05\,T, the increase of the susceptibility below $T_N$ for $H\perp c$ originates from small uncompensated moments in the $ab$ plane. [(b) and (e)] Curie-Weiss fits of the susceptibility between 5 and 20\,K. [(c) and (f)] The TLAF fit of the magnetic susceptibility describes the data accurately. The coefficients are listed in Ref.~\cite{Tamura2002}.}
\end{figure*}

\subsection{Magnetization}
\label{Magn_Results}
Co$^{2+}$ is a magnetic ion with the electronic configuration $d^7$ and octahedral coordination in Na$_2$SrCo(PO$_4$)$_2$. In the high-spin case, spin-orbit coupling (SOC) lifts the degeneracy of the $^4T_1$ dodecet. The lowest Kramers level has effective spin-1/2. The higher-lying Kramers levels have effective spin-3/2 and 5/2. At low temperatures only the low-lying doublet is accessed and an effective spin-1/2 behavior is observed, as confirmed by our analy\-sis of the specific heat. With increasing temperature the higher-lying level with effective spin-3/2 gets populated \cite{Wellm2021}. In the temperature dependence of the inverse susceptibility (Fig.~\ref{Griffith}) the inflection point at around 70\,K marks the increased population of the excited state. The model from Griffith \cite{Griffith1958} for $d^7$ high-spin octahedral complexes describes the experimental data accurately but the theory has also its limitations. In our case, the obtained SOC-constant of 1377(11)\,K is around 2.5 times larger than in Ref.~\cite{Wellm2021}.
The model assumes an ideal octahedral symmetry wherein the splitting of the Co$^{2+}$ multiplet arises from the SOC only. On the other hand, in Na$_2$SrCo(PO$_4$)$_2$ trigonal distortion, as well as the deformation of the octahedra beyond trigonal symmetry, should lead to a crystal-field splitting of the $t_{2g}$ levels (see also Sec.~\ref{ab_initio_res}) and, consequently, to deviations from the simplified model by Griffith \cite{Griffith1958}.\\
\indent We now concentrate on the temperature range below 20\,K where the influence of the excited states should be negligible and the pure spin-1/2 regime can be assumed. The Curie-Weiss fits, $\chi=\chi_0+C/(T-\theta)$, between 5 and 20 K return the Curie-Weiss temperature $\theta_{CW}=-3.07(3)$\,K, Landé factor $g=5.8(3)$, $\chi_0=7.8(5)\cdot10^{-3}$\,emu/mol for $H\parallel c$ and $\theta_{CW}=-1.44(3)$\,K, $g=3.7(2)$, $\chi_0=8.5(3)\cdot10^{-3}$\,emu/mol for $H\perp c$ [Figs.~\ref{Magn}(b) and \ref{Magn}(e)]. In both cases, $g$ values are given for the effective spin-1/2 of the ground-state Kramers doublet. $\chi_0$ was added to account for a Van-Vleck contribution. From $\theta_{CW}=1.5J/k_{\text{B}}$ for nearest-neighbor triangular antiferromagnets, $J_{\text{z}}/k_{\text{B}}=2.05(3)\,\text{K}$ and $J_{\text{xy}}/k_{\text{B}}=0.96(2)\,\text{K}$ are obtained. The average value $J_{\text{av}}/k_{\text{B}}=(2J_{\text{xy}}+J_{\text{z}})/3k_{\text{B}}=1.32(3)\,\text{K}$ matches perfectly the value extracted from the heat capacity data. The large difference of the $g$ values indicates an anisotropy of around 37\%. A similar anisotropy is obtained from saturation magnetization, $M_s=g\mu_{\text{B}}S$, measured at 0.4\,K: $g=5.9(3)$ for $H\parallel c$, $g=3.7(2)$ for $H\perp c$. In contrast, a much smaller anisotropy of 12\% ($g=4.7(3)$ for $H\parallel c$, $g=4.1(2)$ for $H\perp c$) was measured for the Ba compound (Fig.~\ref{MvH}). In the case of an undistorted CoO$_6$ octahedron, one expects the isotropic value of $g=4.33$ assuming the effective spin-1/2. The disparity between $g_{\text{z}}$ and $g_{\text{xy}}$ is rooted in the distortion of the CoO$_6$ octahedra. The relatively weak anisotropy in Na$_2$BaCo(PO$_4)_2$ is caused by the trigonal distortion of only 4.6\% at 10\,K. The larger trigonal distortions of around 5.4\% (Co1O$_6$) and 8.8\% (Co2O$_6$) and the distortions beyond trigonal of around 2.6\% (Co1O$_6$) and 2.9\% (Co2O$_6$) in the Sr compound at 10\,K are the origin for the larger $g$ anisotropy.\\
\indent The transition at 600\,mK seen in the specific heat is due to an antiferromagnetic order, because it appears in the susceptibility too [see Fig.~\ref{Magn}(a) and \ref{Magn}(d)]. The transition is observed in magnetic fields up to 0.1\,T and is suppressed at 1\,T. We note in passing that the magnetization curves measured at 400\,mK show some nonlinearities for both directions of the applied field. These effects may indicate the formation of intermediate phases, such as the 1/3-plateau known from other triangular antiferromagnets, and affect the field evolution of $T_N$. However, measurements at much lower temperatures would be needed to resolve these intermediate phases. Such measurements lie beyond the scope of our present study.\\
\indent On the other hand, our data shed light on the nature of the magnetic order in low fields. In the field of 0.01\,T, we observe a drastically different behavior depending on the direction of the applied field. Below $T_N$, the susceptibility decreases for $H\parallel c$ and abruptly increases for $H\perp c$. This behavior can be compared to the textbook case of an easy-axis antiferromagnet where the susceptibility decreases in the field applied along the easy axis and becomes temperature-independent in the field applied perpendicular to the easy axis. Considering $g_{\text{z}}>g_{\text{xy}}$ and the stronger exchange couplings manifested by the more negative Curie-Weiss temperature for $H\parallel c$, we identify $c$ as the easy direction. Indeed, the evolution of the susceptibility below $T_N$ suggests a large and fully compensated spin component along $c$. Additionally, an uncompensated spin component is expected in the $ab$ plane, similar to Refs. \cite{Gitgeatpong2015, Gitgeatpong2017}. It leads to the abrupt increase in the susceptibility below $T_N$ and further manifests itself by the hysteretic behavior in $M(H)$. The hysteresis is shown in the inset of Fig.~\ref{MvH} and the in-plane moment obtained from the saturation magnetization is $\mu_{\text{Co}^{2+}}=0.066(4)\mu_{\text{B}}/\text{f.u.}$. This is equal to 3.5\% of the out-of-plane saturated moment for $B\perp c$. Overall, our data are consistent with the Y-type spin pattern expected in XXZ triangular antiferromagnets with the large easy-axis anisotropy \cite{Yamamoto2019}. The uncompensated $ab$-component would arise if the Y pattern is slightly turned toward the $ab$ plane. A neutron scattering study would be interesting in order to verify this conjecture.\\
\indent In low-dimensional systems or frustrated antiferromagnets, short-range magnetic order often manifests itself by a broad hump in the magnetic susceptibility \cite{Ranjith2019}. For spin-1/2 TLAFs, the broad hump can be described by a Pad\'{e} approximated high-temperature series expansion with $g$ and the exchange coupling $J$ as adjustable para\-meters \cite{Elstner1993, Tamura2002}.
The corresponding fits of the data below 25\,K collected at 0.1\,T are shown in Figs.~\ref{Magn}(c) and \ref{Magn}(f) and yield the exchange interaction $J_{\text{z}}/k_{\text{B}}=1.76(1)\,\text{K}$ and $g=5.9(2)$ for $H\parallel c$ and $J_{\text{xy}}/k_{\text{B}}=1.04(1)\,\text{K}$ and $g=3.9(1)$ for $H\perp c$. The thus obtained $g$ values match the values from the Curie-Weiss fits and $M(H)$ measurements. $J_{\text{z}}$ and $J_{\text{xy}}$ are in good agreement with the values from the Curie-Weiss fits and the average $J_{\text{av}}/k_{\text{B}}=1.28(1)\,\text{K}$ matches the value obtained from the specific heat data. The good description of the susceptibility data with the XXZ model suggests that any off-diagonal anisotropy terms should be small. Indeed, Ref.~\cite{Gao2022} arrives at a similar conclusion for Na$_2$BaCo(PO$_4$)$_2$ using extensive numerical simulations and comparisons to the experimental data also at temperatures well below $J_{\text{av}}$.

\subsection{\textit{Ab initio} results}
\label{ab_initio_res}
\begin{figure}
\includegraphics{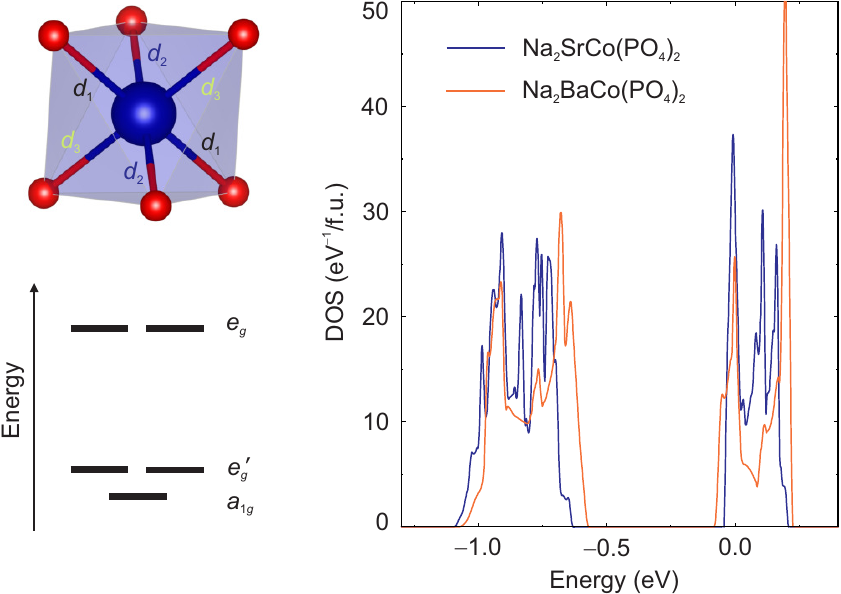}
\caption{\label{fig:DFT}
Local distortions of the CoO$_6$ octahedra, crystal-field splitting of $3d$ orbitals, and the corresponding density of states for Na$_2$SrCo(PO$_4)_2$ and Na$_2$BaCo(PO$_4)_2$. The Fermi level is at zero energy. Note that calculations are performed on the PBE level, resulting in the metallic band structures.
}
\end{figure}
PBE band structure of Na$_2$SrCo(PO$_4)_2$ is metallic because electronic correlations relevant to a Mott insulator are not taken into account. A comparison to Na$_2$BaCo(PO$_4)_2$ (Fig.~\ref{fig:DFT}) shows the nearly unchanged bandwidth that implies a similar strength of exchange couplings in the Sr and Ba compounds. On the other hand, re-structuring of both bands and density of states clearly indicates the increasing distortion of the CoO$_6$ octahedra in the Sr compound.\\
\indent Orbital energies summarized in Table~\ref{tab:orbital} show that trigonal distortion of the CoO$_6$ octahedra leads to the crystal-field splitting of 0.1\,eV within the $t_{2g}$ levels. The lower energy of the $a_{1g}$ orbital corresponds to an easy-axis anisotropy of the $g$ tensor~\cite{liu2021}, in agreement with the experimental $g_{\text{z}}>g_{\text{xy}}$. This splitting is almost unchanged in Na$_2$SrCo(PO$_4)_2$, but an additional splitting between the two $e_g'$ levels appears. Therefore we conclude that the enhanced $g$-tensor anisotropy in Na$_2$SrCo(PO$_4)_2$ should be related not to the trigonal distortion but to the distortion beyond trigonal.\\
\indent Despite the fact that Co1 and Co2 are two distinct positions in the crystal structure of Na$_2$SrCo(PO$_4)_2$, their crystal-field splittings are almost the same, suggesting that experimental $g$ values averaged over Co1 and Co2 should reflect the $g$ values for each of the Co sites. The same is true for the nearest-neighbor exchange couplings evaluated using the mapping procedure. We find $J_{\rm Co1}/k_{\text{B}}=1.5$\,K and $J_{\rm Co2}/k_{\text{B}}=1.8$\,K. Therefore, average exchange coupling extracted from the HTSE fits should be indicative of the interaction strength in each of the triangular planes.
\begin{table}
\caption{\label{tab:orbital}
Orbital energies (in eV) obtained from PBE calculations for Na$_2$SrCo(PO$_4)_2$ and Na$_2$BaCo(PO$_4)_2$. 
}
\begin{ruledtabular}
\begin{tabular}{cr@{\hspace{1cm}}rr}
 \multicolumn{2}{c}{Na$_2$BaCo(PO$_4)_2$} & \multicolumn{2}{c}{Na$_2$SrCo(PO$_4)_2$} \\
          &         & Co1     & Co2\smallskip\\
 $a_{1g}$ & $-0.79$ & $-0.90$ & $-0.89$\smallskip\\
 $e_g'$   & $-0.69$ & $-0.83$ & $-0.82$\smallskip\\
          &         & $-0.80$ & $-0.78$\smallskip\\
 $e_g$    &  0.01   & 0.03    & 0.04\smallskip \\
          &         & 0.08    & 0.07 \\
\end{tabular}
\end{ruledtabular}
\end{table}

\section{Discussion and Summary}
\begin{table}[b]
\caption{Results of the BVS for the SrO and BaO-polyhedra.}
\label{Tab}
\begin{ruledtabular}
\begin{tabular}{lccc}

 & BVS & \parbox{2.49cm}{effective\\ coordination} & \parbox{2.49cm}{average bond length (\AA)}\smallskip \\

 Na$_2$SrCo(PO$_4$)$_2$&2.033 & 9.0 & 2.7255\\
\multicolumn{2}{l}{300\,K, $P2_1/a$} && \\[0.5\normalbaselineskip]
 
 Na$_2$SrCo(PO$_4$)$_2$  & 1.678 & 9.8 & 2.8932\\
\multicolumn{2}{l}{1025\,K, $P\overline{3}m1$} &&\\[0.5\normalbaselineskip]
 
 Na$_2$BaCo(PO$_4$)$_2$  & 2.398 & 10.5 & 2.9141\\
\multicolumn{2}{l}{300\,K, $P\overline{3}m1$} &&\\
\end{tabular}
\end{ruledtabular}
\end{table}
\begin{table}
\caption{\label{tab_Vgl}
Ratio of the N\'{e}el temperature $T_{\text{N}}$ to the average exchange coupling $J_{\text{av}}/k_{\text{B}}$ for several spin-1/2 triangular antiferromagnets. $J_{\text{av}}/k_{\text{B}}=(2J_{\text{xy}}+J_{\text{z}})/3k_\text{B}$ and $J_{\text{av}}/k_{\text{B}}=(J+2J')/3k_\text{B}$ for the Co and Cu compounds, respectively.}
\begin{ruledtabular}
\begin{tabular}{lccc}

 & $T_{\text{N}}$ (K) & $J_{\text{av}}/k_{\text{B}}$ (K)& $k_{\text{B}}T_{\text{N}}/J_{\text{av}}$\\
Na$_2$BaCo(PO$_4$)$_2$ \cite{Bader_unpublished}  & 0.14  & 1.3 	& 0.11\smallskip\\ 
Ba$_3$CoSb$_2$O$_9$ \cite{Kamiya2018}	& 3.8    & 19.19 	& 0.20\smallskip\\
Cs$_2$CuBr$_4$ \cite{Ono2005,Zvyagin2014}			& 1.4    & 9.0  	& 0.16\smallskip\\
Cs$_2$CuCl$_4$ 	\cite{Coldea2001}		& 0.62  & 2.42	  	& 0.26\smallskip\\
Na$_2$SrCo(PO$_4$)$_2$	& 0.60  & 1.32  	& 0.45\smallskip\\

\end{tabular}
\end{ruledtabular}
\end{table}
To explain the measured data and to understand the different properties of Na$_2$SrCo(PO$_4$)$_2$ and Na$_2$BaCo(PO$_4$)$_2$, we need to start by comparing the crystal structures of both compounds. The smaller atomic radii of the Sr$^{2+}$ ions compared to the Ba$^{2+}$ ions in the buffer layers 
lead to a decrease of the interlayer distance and hence the interlayer coupling should increase. The buffer ions not only affect the spacing of the magnetic layers as may be expected, but also have a significant influence on the symmetry. 
To explain the lower mono\-clinic symmetry of the Sr compound compared to the Ba compound, the environments of the divalent ions will be compared by calculating the bond valence sum (BVS). In this case, the BVS connects the oxidation state of the alkaline-earth metal to their bond length to the surrounding O$^{2-}$ ions. The needed empirical constants were taken from \cite{Brown1985}. The resulting BVS, effective coordination, and average bond length are listed in Table~\ref{Tab}. In analogy to the BVS, the effective or mean coordination is obtained by adding the surrounding atoms with a weight depending on their distance. In $P2_1/a$, the smaller Sr ions are coordinated by nine O ions instead of ten as for Ba in $P\overline{3}m1$. To ensure the oxidation state of +2 the average bond length has to decrease. This necessitates the tilt and distortion of the octahedra, resulting in the symmetry lowering. The higher average bond length at elevated temperatures allows for the ten-fold coordination and the undistorted structure can form.\\
\indent An important ramification of the symmetry lowering is the formation of magnetic order with an uncompensated spin component in Na$_2$SrCo(PO$_4$)$_2$. Such uncompensated components would typically arise from DM interactions that are controlled by the inversion symmetry. Indeed, in the trigonal Ba compound, every Co-Co bond carries an inversion center which is removed in the Sr compound for two of the three bonds of the triangle. The lack of the bond inversion symmetry is a necessary condition for DM interactions and therefore only possible in the Sr compound. DM interactions favor non-collinear spin arrangement because energy is gained when spins become tilted relative to each other. The formation of uncompensated moments is a typical manifestation of non-collinear spin structures driven by DM interactions.\\
\indent A further consequence of the symmetry lowering is the enhanced $g$-tensor anisotropy. In the isotropic case (undistorted CoO$_6$ octahedron), $g=4.33$ for both field directions. Parallel and perpendicular components of the $g$ tensor split if distortions are introduced \cite{Abragam1970}.  The increase in the $g$-tensor anisotropy is accompanied by the enhancement of the XXZ anisotropy with the anisotropy parameter $\Delta=J_{\text{z}}/J_{\text{xy}}=2.04\,\text{K}/0.96\,\text{K}=2.13$ estimated from the Curie-Weiss temperatures. In comparison, we determined the anisotropy parameter of the Ba compound to be lower, $\Delta=J_{\text{z}}/J_{\text{xy}}=1.68\,\text{K}/1.11\,\text{K}=1.51$ \cite{Bader_unpublished}, which is in good agreement with the calculated literature values \cite{Gao2022}.\\
\indent Table~\ref{tab_Vgl} lists the ratio $k_{\text{B}}T_{\text{N}}/J_{\text{av}}$ for several spin-1/2 triangular antiferromagnets. Compared to other compounds with $k_{\text{B}}T_{\text{N}}/J_{\text{av}}$ of around 0.2, the Ba compound has the lowest ratio, whereas the Sr compound has the highest ratio. Despite very similar exchange couplings, the N\'{e}el temperature is over four times larger in the Sr compound. We ascribe this difference to three possible effects. First, the smaller size of Sr$^{2+}$ leads to a stronger interlayer coupling and increases $T_{\text{N}}$. Second, the enhanced XXZ anisotropy acts to suppress quantum fluctuations and increase $T_{\text{N}}$. At first glance, the $120$\,\degree order requires an easy-plane anisotropy and would be disfavoured by the easy-axis one, but it can also be compatible with the $J_{\text{z}}/J_{\text{xy}} > 1$ regime if one of the spins aligns with the $c$ axis, resulting in the Y-type order \cite{Yamamoto2019}. Such an order is favored by the easy-axis anisotropy and becomes progressively more stable when $J_{\text{z}}/J_{\text{xy}}$ increases. Finally, the third potentially important effect is the distortion of the triangular spin lattice in monoclinic Na$_2$SrCo(PO$_4$)$_2$, as compared to Na$_2$BaCo(PO$_4$)$_2$ with its higher trigonal symmetry.

\begin{figure}
\includegraphics[angle=0,width=0.48\textwidth]{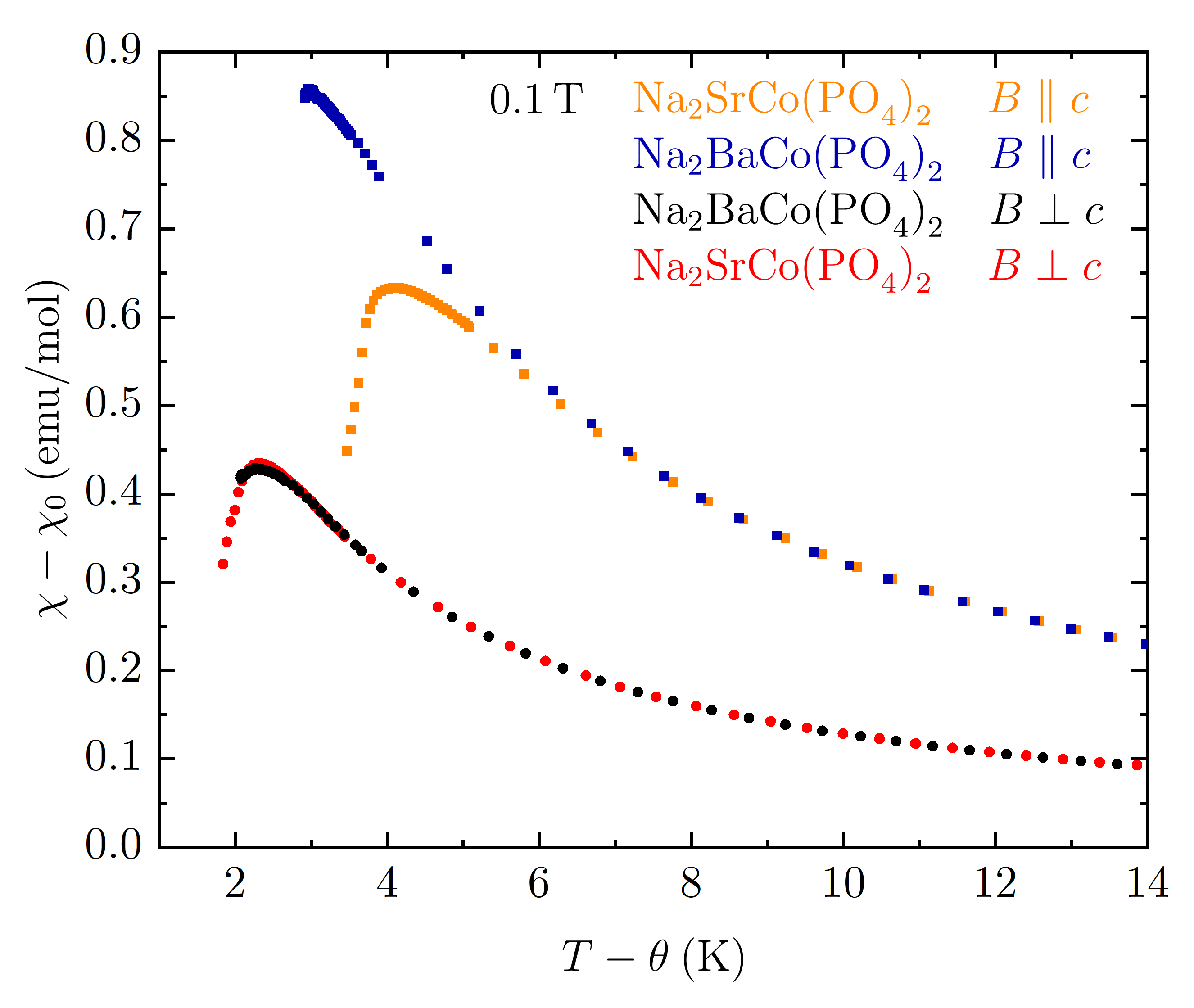}
\caption{\label{SrBaComparison} Comparison of the rescaled susceptibilities of Na$_2$SrCo(PO$_4$)$_2$ and Na$_2$BaCo(PO$_4$)$_2$ \cite{Bader_unpublished} for both field directions at 0.1\,T. Deviations are prominent at low temperatures.}
\end{figure}

This difference between the Sr and Ba compounds can be also visualized by re-scaling magnetic susceptibilities so that they match in the paramagnetic regime (Fig.~\ref{SrBaComparison}). At low temperatures, the largest deviation between the Sr and Ba compounds is seen for $H\parallel c$. It further indicates that easy-axis anisotropy becomes stronger in the Sr compound, and that spins in the magnetically ordered state are primarily directed along $c$. More generally, we see that even a small monoclinic distortion can have a strong influence on the ordering temperature of triangular antiferromagnets. Whereas the ratio $k_{\text{B}}T_{\text{N}}/J_{\text{av}}$ of Na$_2$SrCo(PO$_4$)$_2$ is higher than in other TLAFs, $k_{\text{B}}T_{\text{N}}/J_{\text{av}}$ of Na$_2$BaCo(PO$_4$)$_2$ is lower. One interesting possibility is that a structural distortion is present in the Ba compound too, but contrary to the Sr compound this distortion reduces $T_{\text{N}}$. Detailed studies of the Na$_2$BaCo(PO$_4$)$_2$ crystal structure could shed further light on this scenario.

In this paper, we investigated the compound Na$_2$SrCo(PO$_4$)$_2$, a close analog of the recently studied quantum spin liquid candidate Na$_2$BaCo(PO$_4$)$_2$. It crystallizes in the monoclinic $P2_1/a$ space group. At around 650\,K, the structure transforms into $C2/m$ and further evolves toward trigonal with increasing temperature until the transition is completed above 1025\,K. The change in symmetry compared to Na$_2$BaCo(PO$_4$)$_2$ has several repercussions for the magnetic behavior. DM interactions are now possible and lead to the formation of an uncompensated moment in the $ab$ plane. Distortions of the CoO$_6$ octahedra beyond trigonal, which are possible in the lower symmetry, lead to an increase of the $g$ anisotropy and to a larger XXZ anisotropy. The distortion of the triangular lattice reduces the amount of frustration in the system. These aspects lead together with the increase of the interlayer coupling to a four times higher N\'{e}el temperature in Na$_2$SrCo(PO$_4$)$_2$ compared to Na$_2$BaCo(PO$_4$)$_2$.\\
\indent \textit{Note added:} upon completing this manuscript we became aware of the concurrent work on Na$_2$SrCo(PO$_4)_2$ that reports polycrystalline samples and their thermodynamic properties down to 2\,K~\cite{Zhang2022}. The room-temperature monoclinic ($P2_1/a$) crystal structure reported in that work is in full agreement with our results. 

\acknowledgments
We acknowledge ESRF for providing beamtime for this project, and thank Andy Fitch, Catherine Dejoie, and Ola Grendal for their help during the data collection. AT thanks Mike Zhitomirsky and Oleg Janson for a fruitful discussion. This work was supported by the German Research Foundation (DFG) via Project No. 107745057 (TRR80).

\bibliography{Na2SrCoP2O8.bib}

\end{document}


\begin{titlepage}

\title{\textit{Supplemental Material}\medskip\\Deformation of the triangular spin-1/2 lattice in Na$_2$SrCo(PO$_4$)$_2$}

\author{Vera~P. Bader}
\email{vera.bader@uni-a.de}
\affiliation  {Experimental Physics VI, Center for Electronic Correlations and Magnetism, University of Augsburg, 86159 Augsburg, Germany}

\author{Jan Langmann}
\affiliation  {Chemical Physics and Materials Science , University of Augsburg, 86159 Augsburg, Germany}

\author{Philipp Gegenwart}
\affiliation{Experimental Physics VI, Center for Electronic Correlations and Magnetism, University of Augsburg, 86159 Augsburg, Germany}

\author{Alexander~A. Tsirlin}
\email{altsirlin@gmail.com}
\affiliation{Experimental Physics VI, Center for Electronic Correlations and Magnetism, University of Augsburg, 86159 Augsburg, Germany}
\affiliation{Felix Bloch Institute for Solid-State Physics, University of Leipzig, 04103 Leipzig, Germany}

\begin{figure*}[b]
  \centering
  \includegraphics[width=0.8\textwidth]{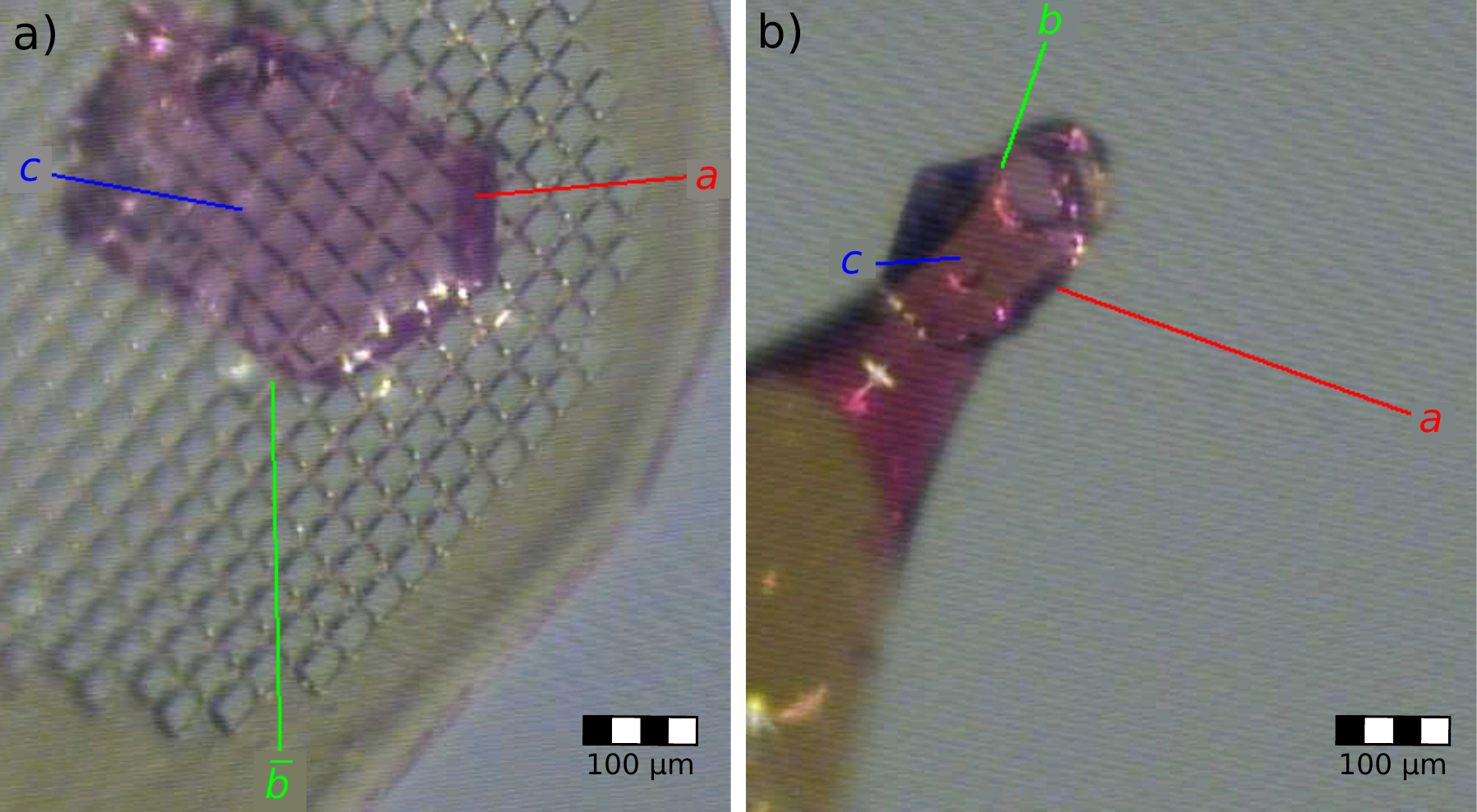}
  \caption{Photographic images of the a)~as-cast (twinned) and b)~cut \Na\ sample (single-domain) used in the room-temperature single-crystal x-ray diffraction study. Crystal axes $a$, $b$ and $c$ are indicated by colored lines and refer to the monoclinic unit cells of a)~the major twin domain and b)~the only present domain.}
  \label{fig:sample-photo}
\end{figure*}

\maketitle
\end{titlepage}

\begin{figure*}
  \centering
  \includegraphics[width=0.8\textwidth]{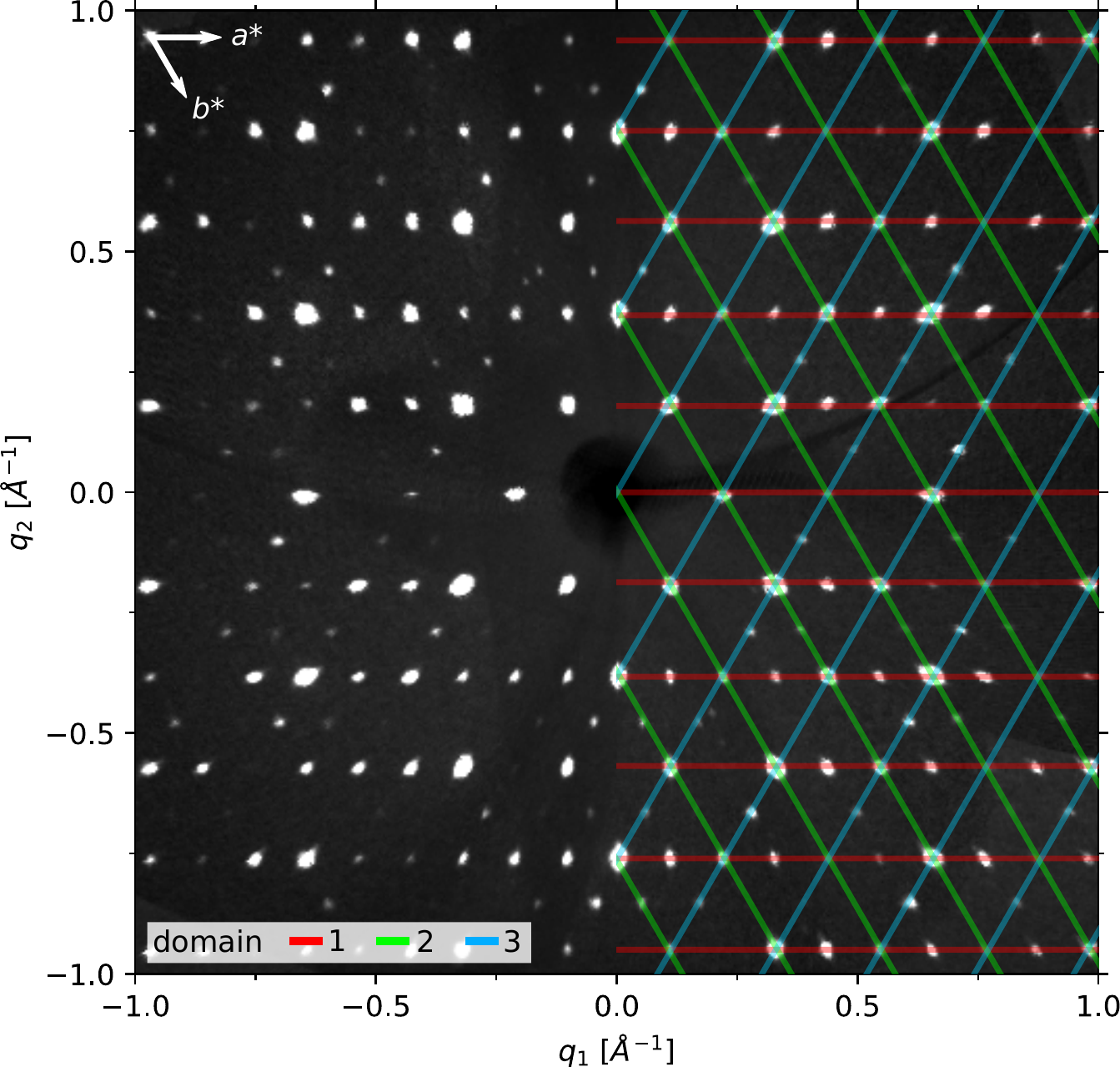}
  \caption{Illustration of the triple twinning in the as-cast \Na\ crystal (Fig.~\ref{fig:sample-photo}a). The shown reconstruction of the reciprocal-space plane $(hk0)$ is based on room-temperature single-crystal x-ray diffraction data and refers to the unit cell of the presumed trigonal high-temperature phase (space group $P\overline{3}m1$; see main paper). Rows of reflections belonging to the different twin domains are indicated by semi-transparent colored lines in the right half of the picture.}
  \label{fig:hk0-twinning}
\end{figure*}

\begin{table*}
  \centering
  \caption{Crystal data and structure refinement as obtained from x-ray diffraction data collected on the as-cast \Na\ single crystal (Fig.~\ref{fig:sample-photo}a) at room temperature.}
  \begin{tabular}{p{0.3\textwidth}>{\centering\arraybackslash}p{0.4\textwidth}}
    \hline
    \multirow{3}{*}{unit cell dimensions} & $a =$~9.1950(5)\,\AA\\
                                          & $b =$~5.2681(3)\,\AA\\
                                          & $c =$~13.5241(8)\,\AA\\
                                          & $\beta =$~90.037(2)\,\degree \\
                                          & $V =$~655.11(6)\,\AA$^3$\\[0.1cm]
    calculated density & 3.8779\,g$\cdot$cm$^{-3}$\\[0.1cm]
    crystal size & 76$\times$306$\times$413\,$\mu$m$^3$\\[0.1cm]
    twin element & $\left(\begin{array}{rrr} 0.5 & 0.4962 & 0\\
                            -1.5116 & 0.5 & 0\\
                            0.0007 & 0.0002 & -1\end{array}\right)$\\[0.1cm]
    twin ratio & 0.875(3) : 0.066(2) : 0.059(3)\\[0.1cm]
    wave length & 0.56087\,\AA\\[0.1cm]
    transm. ratio (max/min) & 0.7884 / 0.2879\\[0.1cm]
    absorption coefficient & 6.043\,mm$^{-1}$\\[0.1cm]
    $F(000)$ & 724\\[0.1cm]
    $\theta$ range & 1\,\degree \ to 34\,\degree \\[0.1cm]
    range in $hkl$ & -17/17, -10/10, -27/26\\[0.1cm]
    total no. reflections & 42967\\[0.1cm]
    independent reflections & 5383 ($R_\mathrm{int} =$~0.0798)\\[0.1cm]
    reflections with $I \geq 3\sigma(I)$ & 4504\\[0.1cm]
    data / parameters & 4504 / 134\\[0.1cm]
    goodness-of-fit on $F^2$ & 1.18\\[0.1cm]
    \multirow{2}{*}{final $R$ indices [$I \geq 3\sigma(I)$]} & $R =$~0.0439\\
                                          & $wR =$~0.1451\\[0.1cm]
    \multirow{2}{*}{$R$ indices (all data)} & $R =$~0.0524\\
                                          & $wR =$~0.1556\\[0.1cm]
    \multirow{3}{*}{extinction correction} & Becker \& Coppens\\
                                          & (isotropic, Type~I,\\
                                          & Gaussian)\\[0.1cm]
    extinction coefficient & 0.65(4)\\[0.1cm]
    largest diff. peak and hole & 6.13 / -3.08\,e$\cdot$\AA$^{-3}$\\
    \hline
  \end{tabular}
  \label{tab:crysdata-ref-1}
\end{table*}

\begin{figure*}
  \centering
  \includegraphics[width=0.8\textwidth]{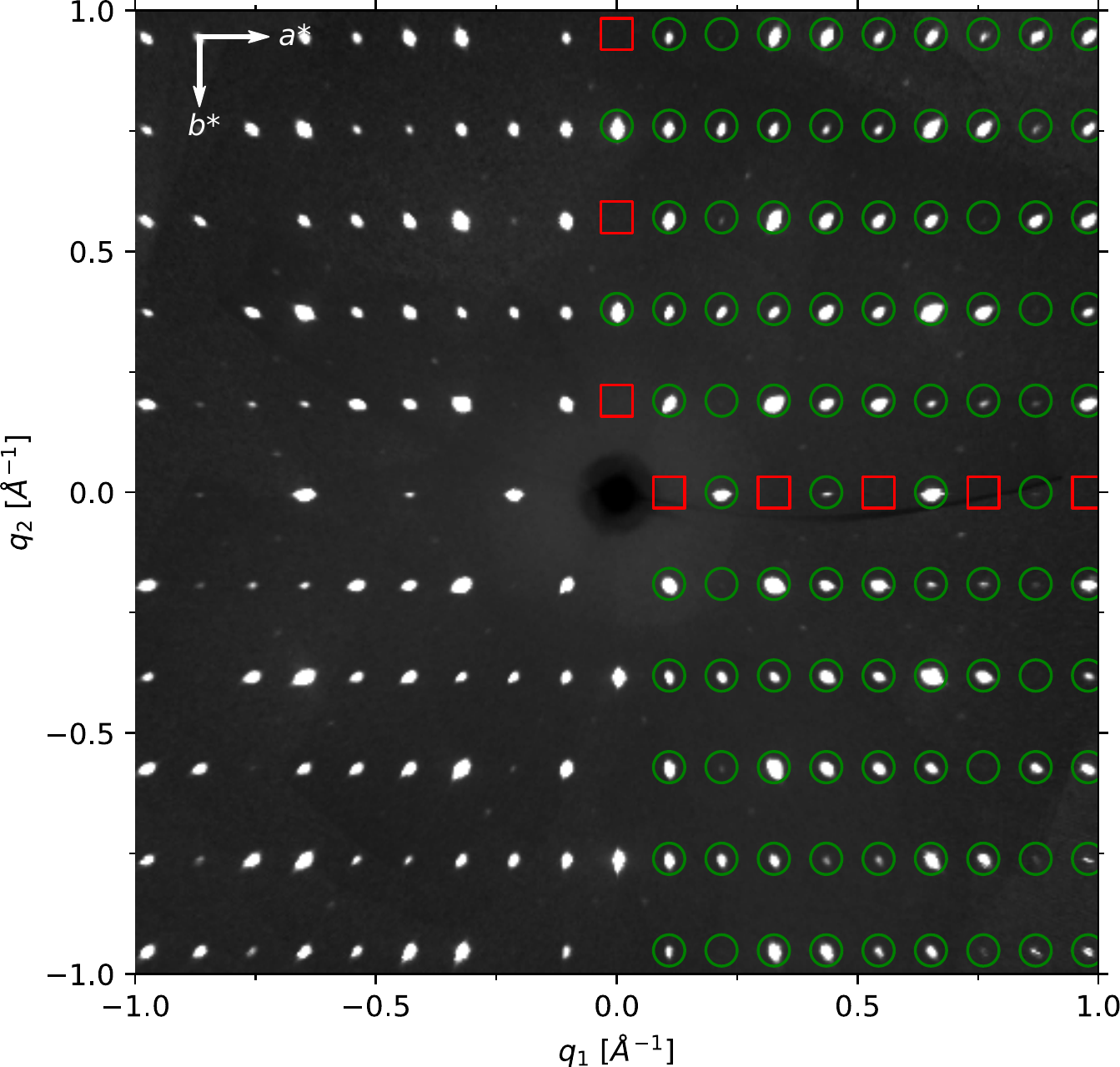}
  \caption{Reconstruction of the reciprocal-space plane $(hk0)$ using room-temperature x-ray diffraction data collected on the cut \Na\ single crystal (Fig.~\ref{fig:sample-photo}b). In the right half of the picture, reflections allowed (green circles) and extinct (red squares) in space group $P 2_1/a$ are highlighted. Note that indicated reciprocal-space axes refer to the monoclinic unit cell of \Na.}
  \label{fig:hk0-plane-ext}
\end{figure*}

\begin{figure*}
  \centering
  \includegraphics[width=0.8\textwidth]{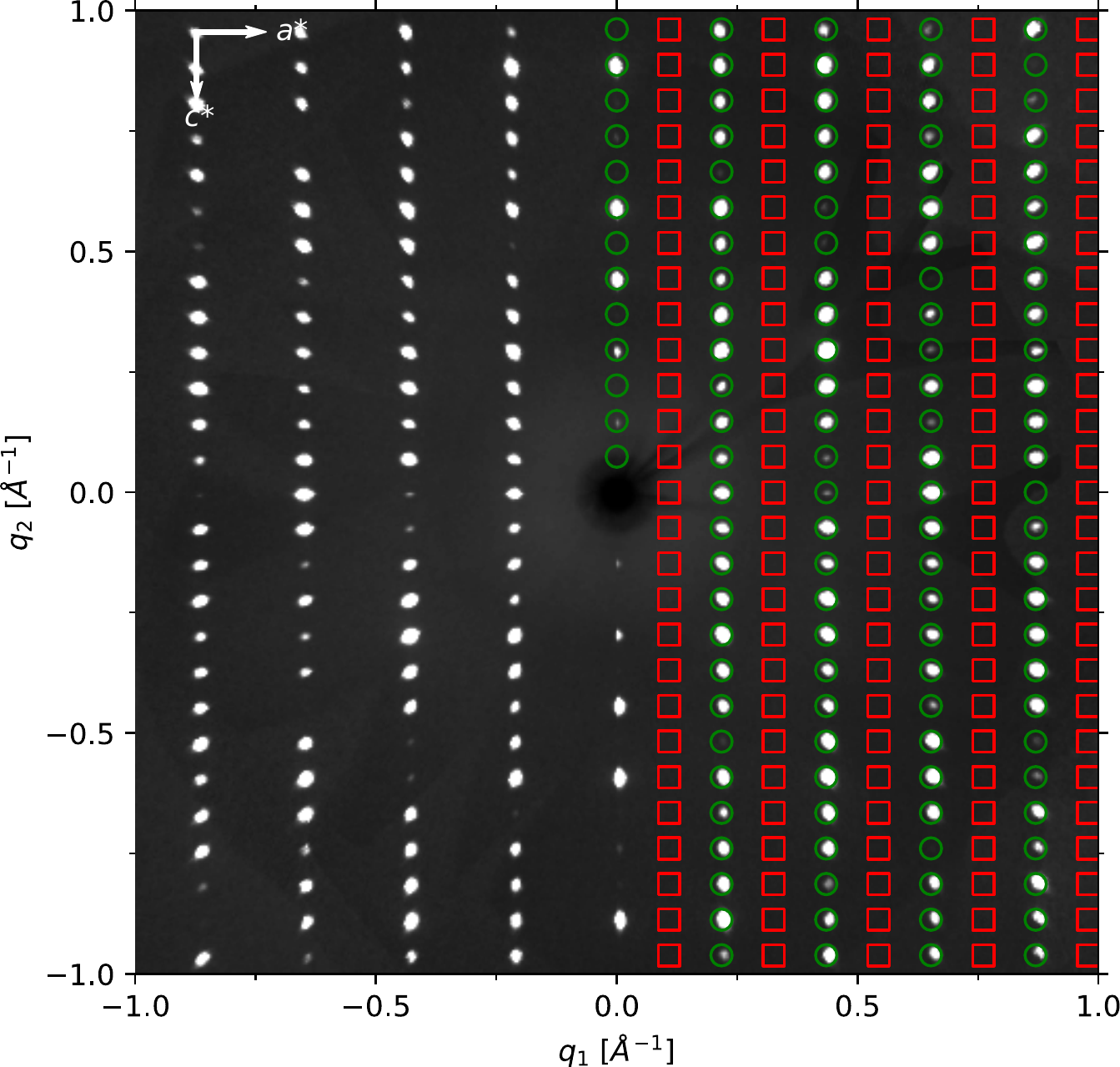}
  \caption{Reconstruction of the reciprocal-space plane $(h0l)$ using room-temperature x-ray diffraction data collected on the cut \Na\ single crystal (Fig.~\ref{fig:sample-photo}b). In the right half of the picture, reflections allowed (green circles) and extinct
(red squares) in space group $P 2_1/a$ are highlighted. Note that indicated reciprocal-space axes refer to the monoclinic unit cell of \Na.}
  \label{fig:h0l-plane-ext}
\end{figure*}

\begin{figure*}
  \centering
  \includegraphics[width=0.8\textwidth]{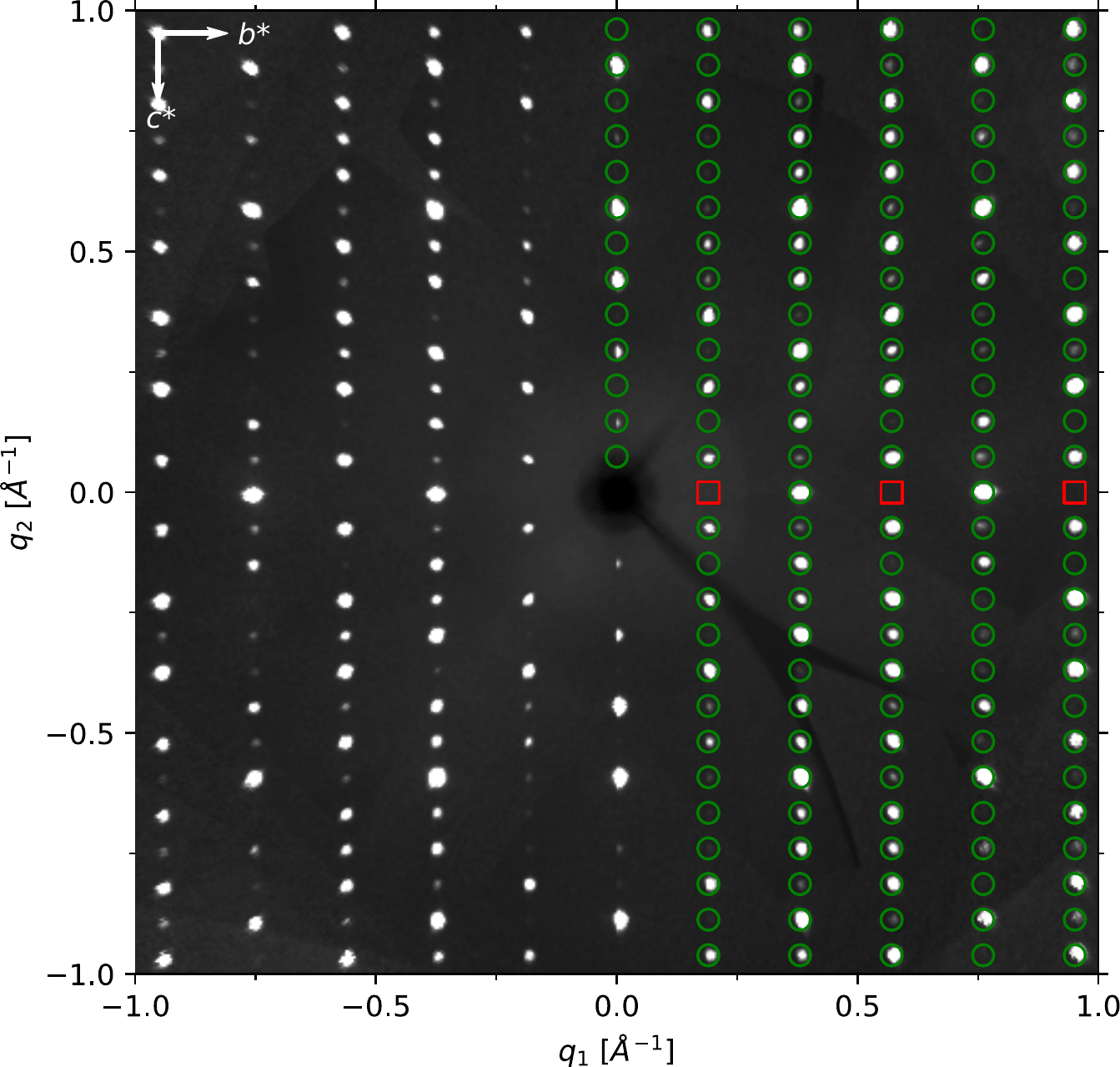}
  \caption{Reconstruction of the reciprocal-space plane $(0kl)$ using
    room-temperature x-ray diffraction data collected on the cut \Na\ single crystal (Fig.~\ref{fig:sample-photo}b). In the right half of the picture, reflections allowed (green circles) and extinct (red squares) in space group $P 2_1/a$ are highlighted. Note that indicated reciprocal-space axes refer to the monoclinic unit cell of \Na.}
  \label{fig:0kl-plane-ext}
\end{figure*}

\begin{table*}
  \centering
   \caption{Crystal data and structure refinement as obtained from x-ray diffraction data collected on the cut \Na\ single crystal (Fig.~\ref{fig:sample-photo}b) at room temperature.}
  \begin{tabular}{p{0.3\textwidth}>{\centering\arraybackslash}p{0.3\textwidth}}
    \hline
    \multirow{3}{*}{unit cell dimensions} & $a =$~9.1931(6)\,\AA\\
                                          & $b =$~5.2586(4)\,\AA\\
                                          & $c =$~13.5269(9)\,\AA\\
                                          & $\beta =$~90.071(3)\,\degree \\
                                          & $V =$~653.93(8)\,\AA$^3$\\[0.1cm]
    calculated density & 3.8849\,g$\cdot$cm$^{-3}$\\[0.1cm]
    crystal size & 76$\times$141$\times$213\,$\mu$m$^3$\\[0.1cm]
    wave length & 0.56087\,\AA\\[0.1cm]
    transm. ratio (max/min) & 0.6773 / 0.4418\\[0.1cm]
    absorption coefficient & 6.054\,mm$^{-1}$\\[0.1cm]
    $F(000)$ & 724\\[0.1cm]
    $\theta$ range & 1\,\degree \ to 34\,\degree \\[0.1cm]
    range in $hkl$ & -18/18, 0/10, 0/27\\[0.1cm]
    total no. reflections & 96409\\[0.1cm]
    independent reflections & 5470 ($R_\mathrm{int} =$~0.0349)\\[0.1cm]
    reflections with $I \geq 3\sigma(I)$ & 5118\\[0.1cm]
    data / parameters & 5470 / 132\\[0.1cm]
    goodness-of-fit on $F^2$ & 1.02\\[0.1cm]
    \multirow{2}{*}{final $R$ indices [$I \geq 3\sigma(I)$]} & $R =$~0.0170\\
                                          & $wR =$~0.0743\\[0.1cm]
    \multirow{2}{*}{$R$ indices (all data)} & $R =$~0.0186\\
                                          & $wR =$~0.0831\\[0.1cm]
    \multirow{3}{*}{extinction correction} & Becker \& Coppens\\
                                          & (isotropic, Type~I,\\
                                          & Gaussian)\\[0.1cm]
    extinction coefficient & 1.22(4)\\[0.1cm]
    largest diff. peak and hole & 1.50 / -1.78\,e$\cdot$\AA$^{-3}$\\
    \hline
  \end{tabular}
  \label{tab:crysdata-ref-2}
\end{table*}

\begin{table*}
  \centering
  \caption{Refined fractional atomic coordinates and mean-square atomic displacement parameters obtained from x-ray diffraction  data collected on the cut \Na\ single crystal (Fig.~\ref{fig:sample-photo}b) at room temperature.}
  \begin{tabular}{>{\centering}p{0.1\textwidth}
    >{\centering}p{0.15\textwidth}>{\centering}p{0.15\textwidth}
    >{\centering}p{0.15\textwidth}>{\centering\arraybackslash}p{0.15\textwidth}}
    & \multicolumn{3}{c}{\textbf{fractional atomic coordinates}} &
    \textbf{\textit{U}\textsubscript{eq}}\\
    \hline
    atom & $x$ & $y$ & $z$ & [\AA$^2$]\\
    \hline
    O1 & 0.70178(9) & 0.50090(16) & 0.92242(6) & 0.01196(15)\\
    O2 & 0.15411(8) & 0.00542(15) & 0.38602(6) & 0.01000(14)\\
    O3 & 0.95015(8) & 0.30437(14) & 0.91005(5) & 0.01049(13)\\
    O4 & 0.41815(7) & 0.72734(13) & 0.88467(5) & 0.00926(12)\\
    O5 & 0.39957(8) & 0.80737(15) & 0.41800(6) & 0.01240(14)\\
    O6 & 0.87226(8) & 0.21962(15) & 0.41167(6) & 0.01233(14)\\
    O7 & 0.82353(10) & 0.45621(15) & 0.75819(6) & 0.01199(16)\\
    O8 & 0.35195(10) & 0.02691(16) & 0.25681(6) & 0.01270(16)\\
    P1 & 0.84842(3) & 0.50923(4) & 0.867019(18) & 0.00515(5)\\
    P2 & 0.31923(3) & 0.02805(4) & 0.366565(18) & 0.00533(5)\\
    Na1 & 0.81911(7) & 0.48619(11) & 0.08944(4) & 0.01392(12)\\
    Na2 & 0.34876(7) & 0.01638(11) & 0.58943(4) & 0.01403(12)\\
    Co1 & 0 & 0 & 0 & 0.00612(4)\\
    Co2 & 0.5 & 0.5 & 0.5 & 0.00621(4)\\
    Sr1 & 0.035333(9) & 0.042901(17) & 0.750259(5) & 0.00883(3)\\
    \hline
  \end{tabular}
  \label{tab:struct-model}
\end{table*}

\begin{table*}
  \centering
  \caption{Refined anisotropic mean-square atomic displacement parameters obtained from x-ray diffraction data collected on the cut \Na\ single crystal (Fig.~\ref{fig:sample-photo}b) at room temperature.}
  \begin{tabular}{>{\centering}p{0.07\textwidth}
    >{\centering}p{0.12\textwidth}>{\centering}p{0.12\textwidth}
    >{\centering}p{0.12\textwidth}>{\centering}p{0.15\textwidth}
    >{\centering}p{0.15\textwidth}>{\centering\arraybackslash}p{0.15\textwidth}}
    & \multicolumn{6}{c}{\textbf{mean-square atomic displacement
                            parameters} [\AA$^2$]}\\
    \hline
    atom & $U_{11}$ & $U_{22}$ & $U_{33}$ & $U_{12}$ & $U_{13}$ & $U_{23}$\\
    \hline
    O1 & 0.0069(2) & 0.0163(3) & 0.0127(3) & -0.0009(2) & 0.0033(2) & 0.0001(2)\\
    O2 & 0.0054(2) & 0.0160(3) & 0.0086(2) & -0.00127(19) & 0.00084(18) & -0.00138(19)\\
    O3 & 0.0112(2) & 0.0080(2) & 0.0123(2) & 0.00255(19) & -0.00086(18) & 0.00310(19)\\
    O4 & 0.0118(2) & 0.0067(2) & 0.0093(2) & 0.00306(18) & -0.00046(17) & 0.00078(17)\\
    O5 & 0.0135(2) & 0.0102(2) & 0.0135(2) & 0.0042(2) & -0.0010(2) & 0.0038(2)\\
    O6 & 0.0125(2) & 0.0092(2) & 0.0153(3) & 0.0029(2) & -0.0039(2) & 0.0035(2)\\
    O7 & 0.0140(3) & 0.0158(3) & 0.0062(2) & -0.0011(2) & -0.0020(2) & -0.00291(19)\\
    O8 & 0.0136(3) & 0.0177(3) & 0.0069(2) & 0.0012(2) & 0.0024(2) & 0.0001(2)\\
    P1 & 0.00520(8) & 0.00508(8) & 0.00517(8) & -0.00024(6) & -0.00023(6) & -0.00010(5)\\
    P2 & 0.00483(8) & 0.00564(8) & 0.00552(9) & 0.00006(5) & -0.00004(6) & -0.00026(5)\\
    Na1 & 0.0144(2) & 0.01656(20) & 0.0107(2) & -0.00545(16) & -0.00161(16) & 0.00124(14)\\
    Na2 & 0.0143(2) & 0.0168(2) & 0.0109(2) & -0.00497(16) & 0.00135(16) & -0.00128(14)\\
    Co1 & 0.00633(7) & 0.00616(6) & 0.00588(7) & 0.00009(4) & 0.00003(5) & 0.00047(4)\\
    Co2 & 0.00623(7) & 0.00655(6) & 0.00585(7) & -0.00058(4) & -0.00032(5) & -0.00024(4)\\
    Sr1 & 0.00946(5) & 0.01106(5) & 0.00598(4) & -0.00115(2) & -0.00013(2) & 0.000382(18)\\
    \hline
  \end{tabular}
  \label{tab:struct-model-ADPs}
\end{table*}

\begin{figure*}
  \centering
  \includegraphics[width=0.8\textwidth]{Refinement800K}
  \caption{Rietveld refinement for the synchrotron XRD data collected at 800 K in the monoclinic space group $C2/m$. The black dots are the experimental data, the red line is the Rietveld fit, the blue line is the difference between experiment and fit, and the black ticks mark the peak positions.}
  \label{fig:Refinement800K}
\end{figure*}

\begin{table*}
  \centering
  \caption{Refined fractional atomic coordinates and mean-square atomic displacement parameters obtained from synchrotron x-ray powder diffraction data collected at 800\,K ($C2/m$). The cell parameters are $a=9.22467(2)\,\r{A}$, $b=5.32301(1)\,\r{A}$, $c=13.73148(3)\,\r{A}$, and $\beta=90.0717(2)$\,\degree .}
  \begin{tabular}{>{\centering}p{0.1\textwidth}
    >{\centering}p{0.15\textwidth}>{\centering}p{0.15\textwidth}
    >{\centering}p{0.15\textwidth}>{\centering\arraybackslash}p{0.15\textwidth}}
    & \multicolumn{3}{c}{\textbf{fractional atomic coordinates}} &
    \textbf{\textit{U}\textsubscript{eq}}\\
    \hline
    atom & $x$ & $y$ & $z$ & [\AA$^2$]\\
    \hline
 O1   & 0.6960(10) & 0.5 & 0.9204(7) & 0.034(3) \\
 O2   & 0.1619(8) &  0 & 0.3915(5) & 0.009(3)  \\
 O3   & 0.9350(6) &  0.278(1) & 0.8994(5)& 0.025(2) \\
 O4   & 0.4037(7) & 0.773(1) & 0.4121(5) & 0.035(2) \\
 O5   & 0.8257(8) & 0.5 & 0.7562(8) & 0.026(4)\\
 O6   & 0.3480(8) &  0 & 0.2582(8) & 0.034(4)\\
 P1   & 0.8435(4) & 0.5 & 0.8675(4) &  0.012(1)\\
 P2   & 0.3277(4) & 0 & 0.3671(4) & 0.013(1) \\
 Na1  & 0.8283(6) & 0.5 & 0.0882(5) & 0.033(2) \\
 Na2  & 0.3525(6) & 0 & 0.5927(5) & 0.040(2) \\
 Co1  & 0 & 0 & 0 & 0.0198(8)\\
 Co2  & 0.5 & 0.5 & 0.5 & 0.0187(9) \\
 Sr1  & 0.0212(1) & 0 & 0.7498(2) & 0.0257(3)\\
\hline
  \end{tabular}
  \label{tab:struct-800K}
\end{table*}

\begin{table*}
  \centering
  \caption{Refined fractional atomic coordinates and mean-square atomic displacement parameters obtained from synchrotron x-ray powder diffraction data collected at 1025\,K ($P\overline{3}m1$). The cell parameters are $a=5.338296(8)\,\r{A}$ and $c=6.90823(1)\,\r{A}$.}
  \begin{tabular}{>{\centering}p{0.1\textwidth}
    >{\centering}p{0.15\textwidth}>{\centering}p{0.15\textwidth}
    >{\centering}p{0.15\textwidth}>{\centering\arraybackslash}p{0.15\textwidth}}
    & \multicolumn{3}{c}{\textbf{fractional atomic coordinates}} &
    \textbf{\textit{U}\textsubscript{eq}}\\
    \hline
    atom & $x$ & $y$ & $z$ & [\AA$^2$]\\
    \hline
O1   & 0.1798(3) & 0.8202(3) & 0.3084(4) &  0.0356(10) \\
O2   & 1/3 &  2/3 & 0.0169(9)  & 0.040(2)  \\
P1   & 1/3 & 2/3 & 0.2349(4) & 0.0144(5) \\
Na1  & 1/3 & 2/3 & 0.6794(5) & 0.044(1) \\
Co1  & 0 & 0 & 0.5 & 0.0225(6)\\
Sr1  & 0 & 0 & 0 & 0.0310(5)\\
\hline
  \end{tabular}
  \label{tab:struct-1025K}
\end{table*}

\begin{figure*}
  \centering
  \includegraphics[width=0.66\textwidth]{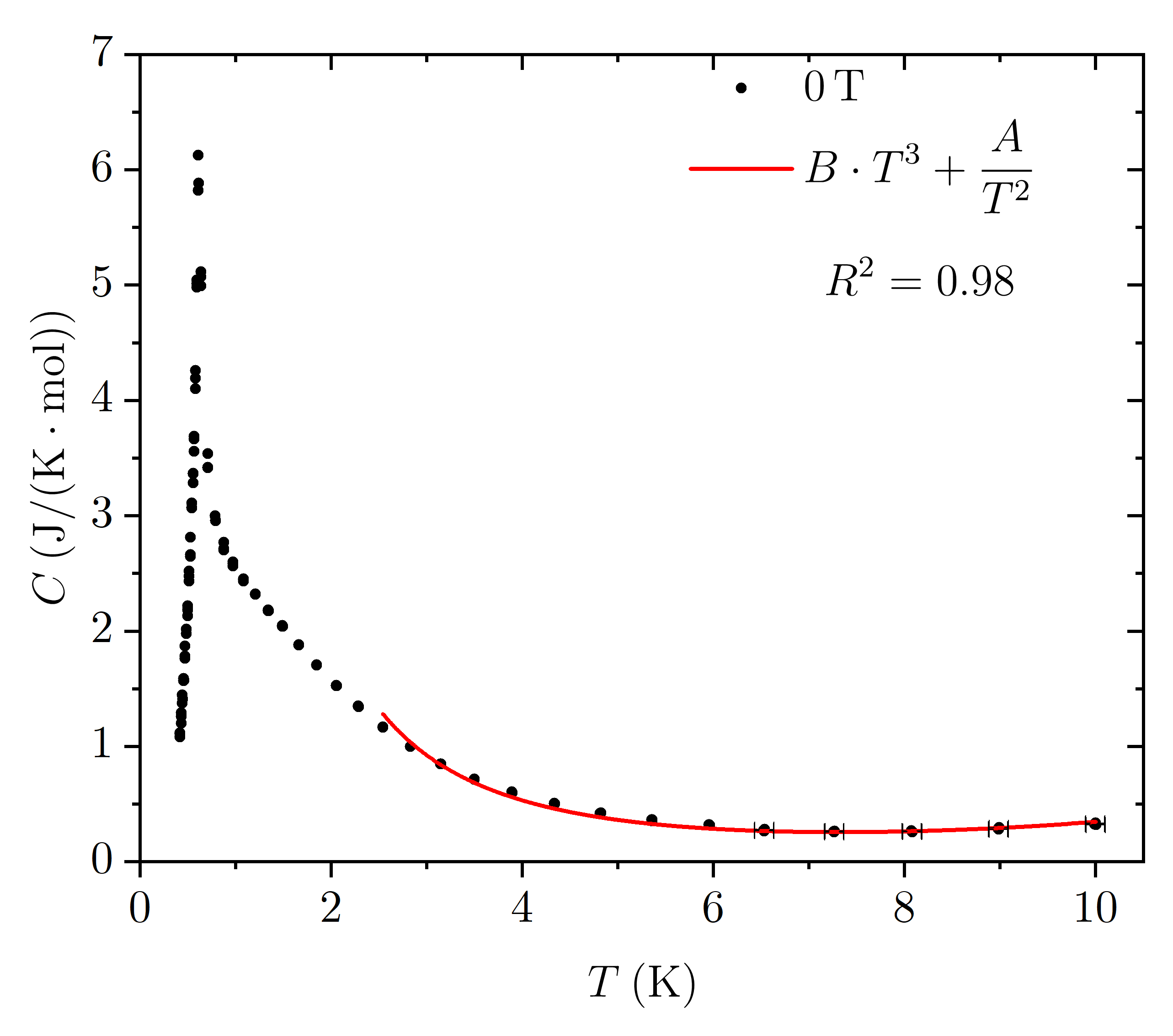}
  \caption{The zero-field heat capacity data between 2.5\,K and 10\,K were fitted to $B\cdot T^3 + A/T^2$, where the first term is the phonon part and the second one is the lowest-order term in the high-temperature series expansion for the magnetic specific heat.}
\end{figure*}